\documentclass{jpp}
\usepackage{graphicx}

\usepackage[utf8]{inputenc}
\usepackage[T1]{fontenc}
\usepackage{amsmath}
\usepackage{xspace}
\usepackage{xcolor}
\usepackage{hyperref}
\usepackage{rotating}

\newcommand{\gkyl}{\texttt{Gkeyll}\xspace}

\newcommand{\uV}{\boldsymbol{u}} 
 
\newcommand{\BV}{\boldsymbol{B}} 
 
\newcommand{\EV}{\boldsymbol{E}} 
 
\newcommand{\xV}{\boldsymbol{x}} 
\newcommand{\vV}{\boldsymbol{v}} 
 
\newcommand{\tsti}{t_{\text{str},i}}
\newcommand{\vta}{v_{\text{th},\alpha}}
\newcommand{\vte}{v_{\text{th},e}}
\newcommand{\vti}{v_{\text{th},i}}
\newcommand{\tturn}{t_{\text{turn}}}

\shorttitle{From Weibel seeds to collisionless dynamos beyond pair-plasmas}
\shortauthor{L.~Hanebring, J.~Juno, A.~Hakim, J.~M.~TenBarge, I.~Pusztai}

\title{From Weibel seeds to dynamo beyond pair-plasmas}

\author{Lise Hanebring\aff{1}
  \corresp{\email{liseha@chalmers.se}},
  James Juno\aff{2},
  Ammar Hakim\aff{2},
  Jason M. TenBarge\aff{2}
 \and Istv\'{a}n Pusztai\aff{1}}

\affiliation{\aff{1}Department of Physics and Astronomy, Chalmers University of Technology, Gothenburg 41296, Sweden
\aff{2}Princeton Plasma Physics Laboratory, Princeton, New Jersey 08543, USA}

\begin{document}

\maketitle

\begin{abstract}
Bridging the spatiotemporal scales of magnetic seed field generation and subsequent dynamo amplification in the weakly collisional intracluster medium presents an extreme numerical challenge. We perform collisionless turbulence simulations with initially unmagnetized electrons that capture both magnetic seed generation via the electron Weibel instability and the ensuing dynamo amplification. Going beyond existing pair-plasma studies, we use an ion-to-electron mass ratio of $100$ for which we find electron and ion dynamics are sufficiently decoupled. These simulations are enabled by the 10-moment collisionless fluid solver of \gkyl, which evolves the full pressure tensor for all species. The electron heat-flux closure regulates pressure isotropization and effectively sets the magnetic Reynolds number. We investigate how the strength of the closure influences the transition between a regime reminiscent of previous kinetic pair-plasma simulations and a regime exhibiting dynamo behavior qualitatively similar to magnetohydrodynamics.
\end{abstract}

\section{Introduction}
\label{sec:intro}
The magnetic fields permeating the universe \citep{Carilli2002,Govoni_2004,Brandenburg_2005,Kulsrud_2008,Brandenburg_2023} profoundly affect the development of astrophysical systems from stellar to cosmological scales. They play important roles in the dynamics of astrophysical plasmas, through, e.g., setting the transport properties of the medium, such as the viscosity \citep{Zhuravleva_2019,Kunz_2014} and heat conductivity \citep{RobergClark_2016,RobergClark_2018,Komarov_2016,Komarov_2018}, and by providing channels for bulk plasma heating and particle energization \citep{Sironi_2025,Guo_2024,Amano_2022,Matthews_2020}. While the origin of cosmic magnetic fields remains an open question, growing observational evidence based on Faraday rotation \citep{Bonafede_2010}, synchrotron emission \citep{Feretti_2012}, and Zeeman splitting \citep{Crutcher_2012} indicates dynamically important magnetic fields in galaxies and galaxy clusters. This observational evidence is consistent with these fields being created and sustained through dynamo action. This inherently three-dimensional (3D) process that naturally arises in turbulent plasma environments is believed to have exponentially amplified seed magnetic fields to equipartition levels, where the magnetic energy content is comparable to the kinetic energy of the turbulent flows, on timescales shorter than the Hubble time.     

More than a century after Joseph Larmor proposed the dynamo phenomenon to explain the solar magnetic field \citep{Larmor_1919}, the role of dynamo in producing stellar and planetary magnetic fields is well established through a magnetohydrodynamic (MHD) modeling of these strongly collisional systems \citep{Kapyla_2023,Schubert_2011,Tobias_2021}. The mean free path of Coulomb collisions in galaxies and the intracluster medium (ICM) of galaxy clusters, however, is typically larger than the relevant turbulent length-scales. Thus, the large-scale properties of these weakly collisional environments are governed by kinetic microphysics instead of Coulomb collisions. Even though a large body of research into the turbulent dynamo in such systems is based on MHD modeling \citep{Schekochihin_2004,Porter_2015,Vazza_2017,KorpiLagg_2024}, it has become clear that addressing the question with tools beyond MHD would be desirable \citep{Rincon_2019}. On the other hand, the multi-scale and inherently 3D nature of turbulent dynamo, which is difficult enough to model with MHD, now becomes an outstanding computational challenge. 

A fully kinetic treatment of systems beyond unity mass-ratio of the plasma particle species remains computationally prohibitively expensive. Hybrid (kinetic ion, fluid electron) modeling established the importance of a rich microphysics \citep{Rincon_2016,StOnge_2018}, where instabilities play a key role not only in setting the macroscopic transport properties of the plasma but also breaking adiabatic invariance, which would otherwise strictly inhibit dynamo action \citep{Helander_2016}\footnote{This interplay is why dynamo action is inhibited in the CGL closure \citep{CGL_1956}, which is also not suited to capture the non-gyrotropic physics of the Weibel instability considered here.}. This is an important aspect, since the intra-cluster medium is extremely well magnetized, with typical Larmor radii being of planetary size, compared to the megaparsec scale system size. Another issue is that most of the proposed seed generation mechanisms, such as the Biermann battery \citep{Biermann_1950} at the ionization fronts of the early universe \citep{Subramanian_1994}, yield tiny magnetic fields (typically $\sim 10^{-24} \,\rm T$) \citep{Langer_2018,Subramanian_2016,Durrive_2015}. These fields are unable to magnetize the plasma, in which case the dynamo is significantly less efficient due to a kinetic damping of the magnetic field \citep{pusztaiPRL}. 

Recent fully kinetic studies in pair-plasmas offer a new paradigm where the electron Weibel instability \citep{Weibel_1959}, driven by pressure anisotropies naturally developing in collisionless turbulent plasmas, provides the seed for dynamo \citep{Zhou_2024,Pucci_2021,Sironi_2023}. This instability saturates as the electrons start to get magnetized at the scale of the instability\footnote{This saturation condition borne out of simulations of \citet{Califano_1998} 
can provide a rough estimate, however at small pressure anisotropy relevant to our investigation, saturation occurs when the current in the Weibel filaments reach the \emph{Alfv\'{e}n current} \citep{Kato_2005}.}, 
corresponding to sizable seed fields ($\sim 10^{-14} \,\rm T$ for typical ICM turbulence parameters). Numerical simulations that connect the Weibel seed generation and the dynamo processes are important to establish that fields at large scales, i.e., the outer scales of the turbulence, can also be generated from the Weibel seeds in a self-consistent manner, especially in the light of recent work by \citet{Liu_2025} pointing out that such an inverse cascade could be suppressed by the firehose instability. In addition moving beyond pair-plasmas is desirable to explore the impact of the scale separations induced by the ion-electron mass-ratio. 

In this paper we use the 10-moment solver of \gkyl\footnote{\url{https://github.com/ammarhakim/gkeyll}} \citep{Wang_2015,Wang_2020} to perform 3D simulations of forced turbulence at a mass-ratio where ion and electron processes are significantly decoupled, which incorporate both the electron Weibel instability and the dynamo phases of magnetic field generation. The 10-moment solver of \gkyl evolves the full pressure tensor, thereby retaining relevant pressure anisotropy driven instabilities of high-beta plasmas, including the Weibel instability in the non-magnetized case. In addition, being implicit permits stepping over unimportant spatiotemporal scales, such as the Debye scale, allowing a larger flexibility in exploring various plasma parameter regimes. When the electron pressure isotropization by the heat flux closure is weak, our simulation results are qualitatively similar to previous kinetic simulations in pair plasmas, apart from electrons and ions playing separate roles here. When the electron pressure is more strongly driven towards isotropy, the results align better with expectations from magnetohydrodynamics.    

The article is organized as follows: After describing the numerical model in Sec.~\ref{sec:tenmom}, and defining effective Reynolds numbers in the absence of explicit viscosity and resistivity in Sec.~\ref{sec:reynolds}, we describe the simulation settings in Sec.~\ref{sec:settings} and briefly comment on Weibel instabilities in our context in Sec.~\ref{sec:weibel}. The results of our baseline case are presented in Sec.~\ref{sec:baseline}, then we discuss a scan of electron pressure isotropization in Sec.~\ref{sec:k0escan}. Finally we summarize and discuss our results in Sec.~\ref{sec:conclusions}. We provide additional details on the Weibel instability in the 10-moment fluid model in Appendix~\ref{app:weibel}, and discuss the evolution of pressure anisotropy in our simulations in Appendix~\ref{brazil}.
 
\section{Methods}\label{sec:methods}

\subsection{The 10-moment collisionless fluid system}
\label{sec:tenmom}
Our simulations use the 10-moment collisionless fluid model of the \gkyl continuum plasma physics solver framework. 
The 10-moment model employed here uses the following set of non-relativistic moment equations: 
\begin{align}
    \label{eq:ContinuityEq}
    \frac{\partial n}{\partial t} + \frac{\partial}{\partial x_j} (n u_j) &= 0, \\
    \label{eq:MomentumEq}
    m n \frac{\partial u_j}{\partial t} + m n u_k \frac{\partial u_j}{\partial x_k} + \frac{\partial p_{jk}}{\partial x_k} &= n q \big[ E_j + \epsilon_{jkl} u_k B_l \big], \\
    \label{eq:EnergyEquation}
    \frac{\partial p_{jk}}{\partial t} + \frac{\partial}{\partial x_l} (p_{jk} u_l) + p_{jl} \frac{\partial u_k}{\partial x_l} + p_{kl} \frac{\partial u_j}{\partial x_l} + \frac{\partial q_{jkl}}{\partial x_l} &= \frac{q}{m} \big[ \epsilon_{jlm} p_{lk} B_m + \epsilon_{klm} p_{lj} B_m \big],
\end{align}
where we have suppressed the species index $\alpha$, corresponding to mass and charge $m_\alpha$ and $q_\alpha$, but otherwise $n_\alpha=\int f_\alpha d^3v$ is the density, where $f_\alpha(\xV,\vV,t)$ is the particle distribution function, and the fluid flow velocity vector, pressure tensor, and heat flux tensor components are defined by $u_{j,\alpha}=n_\alpha^{-1} \int  v_j f_\alpha d^3v$, $p_{jk,\alpha}=m_\alpha \int  (v_j-u_j)(v_k-u_k) f_\alpha d^3v$, and $q_{jkl,\alpha}=m_\alpha \int  (v_j-u_j)(v_k-u_k)(v_l-u_l) f_\alpha d^3v$, respectively. Furthermore $E_j$ and $B_j$ denote the electric and magnetic field components, which are evolved according to Maxwell's equations, $\partial_t \BV=-\nabla\times \EV$ and $\partial_t \EV=c^2\nabla\times \BV-\epsilon_0^{-1}\sum_\alpha q_\alpha n_\alpha \uV_\alpha$. The solver uses a locally implicit scheme for the source terms, while solving the homogeneous portion of the equations by the wave propagation method \citep{LeVeque_1997}. This approach allows certain unimportant scales to be under-resolved without leading to numerical difficulties. In particular, our simulations strongly under-resolve the Debye scale, which would not be possible in most particle-in-cell kinetic codes. This provides us greater flexibility in the plasma parameter regimes considered.  

Importantly, the full pressure tensor is evolved for each species, which allows some pressure anisotropy driven instabilities \citep{Gary_1993} to be retained, such as the firehose instability \citep{Walters_2024}, and with particular relevance to the current work, the Weibel instability. Previous work by \citet{Kuldinow_2025} employing the 10-moment system with a zero heat flux closure ($q_{jkl}=0$) demonstrated that the model can predict electron Weibel instability growth rates in the long wavelength and the strongly driven limits and accurately reproduces the high wavenumber cutoff, as well as the nonlinear saturation level, of the instability.      

Here, we use a closure that drives the system towards an isotropic pressure \citep{Wang_2015} 
\begin{equation}
       \frac{\partial q_{jkl,\alpha}}{\partial x_l} = k_{0,\alpha} \frac{\vta}{\sqrt{2}} (p_{jk,\alpha} - p_\alpha \delta_{jk}), 
       \label{closure}
\end{equation}
where $p_\alpha=\delta_{lm}p_{lm,\alpha}/3$ is the scalar pressure and $\vta=\sqrt{2 p_\alpha/(n_\alpha m_\alpha)}$ is the thermal speed. The parameters $k_{0,\alpha}$ adjust the strength of the isotropization for each species. 
This isotropization closure is not derived to capture a specific kinetic process (in contrast to, e.g., the Hammett--Perkins closure \citep{HammettPerkins} for Landau damping), and the parameters $k_{0,\alpha}$ should therefore be regarded as free. However, by studying the parameters' effect in a controlled manner, they can be used to approximate selected aspects of kinetic physics. In particular, as discussed in Sec.~\ref{sec:reynolds}, they regulate the damping of flows and magnetic fields in the unmagnetized limit, allowing one to emulate effective kinetic dissipation. In addition, as we show in Appendix~\ref{app:weibel}, the closure parameter can also be tuned to reproduce key characteristics of the electron Weibel instability. While other closures for the heat flux are available in \gkyl, such as nonlocal \citep{Ng_2017} or gradient-based \citep{Ng_2020}, we choose the local relaxation closure for its explicit connection to dissipation and our ability to determine effective Reynolds numbers from the choice of the closure parameter.

\subsection{The effect of closure parameters and effective Reynolds numbers}
\label{sec:reynolds}

Unlike in resistive-MHD, where the concepts of dynamo theory have mostly been developed, in the 10-moment system, or a collisionless kinetic system for that matter, there is no explicit kinematic viscosity $\nu$ and magnetic diffusivity $\eta$. In addition, the corresponding damping rates of flow and magnetic perturbations can in general have different wavenumber ($k$) dependence to the $\propto k^2$ diffusive behavior. Thus the usual definitions of the fluid and magnetic Reynolds numbers, ${\rm Re}=u_0 L/\nu$ and ${\rm Rm}=u_0 L/\eta$, respectively, are less trivial to evaluate\footnote{In particle based modeling, the effective collisionality may be calculated based on the statistics of pitch angle scattering caused by collective fields \citep{Zhou_2024}.}, and they have different implications concerning the scale separation between the outer scale and the fluctuation cutoff scales. 

To put our results in context, it is nevertheless useful to quantify the magnetic and fluid damping in terms of analogous dimensionless quantities, comparing inverse timescales of processes creating and dissipating magnetic and fluid kinetic energy on the outer scale $L$. The fluid Reynolds number compares inertial processes $\sim u_0/L$ to (viscous) flow damping $\sim \nu/L^2$, while the magnetic Reynolds number compares field line stretching by the flow $\sim u_0/L$ to resistive dissipation of magnetic fields $\sim \eta/L^2$. Thus, a plausible generalization of these quantities is
\begin{equation}
{\rm Re} =  \frac{u_0/L}{\left.\gamma_U^-\right|_{k_0}} \qquad \text{and} \qquad {\rm Rm} =  \frac{u_0/L}{\left.\gamma_B^-\right|_{k_0}},
\end{equation}
where $\left.\gamma_U^-\right|_{k_0}$ and $\left.\gamma_B^-\right|_{k_0}$ are the damping rates of the mass flow and the magnetic field, respectively, taken at the wavenumber corresponding to the box scale $k=k_0\equiv 2\pi/L$.  

In the following, we will show that in the 10-moment system employing the heat flux closure given in Eq.~(\ref{closure}), the damping of the mass flow is governed by the ion closure parameter $k_{0,i}$, and the magnetic field damping by the electron closure parameter $k_{0,e}$. Specifically, the damping rate of the mass flow scales approximately as $\gamma_U^-\propto k^2/k_{0,i}$, that is, with the same $k$-dependence as the diffusive flow damping in a viscous fluid. However, the damping rate of the magnetic field scales approximately as $\gamma_B^-\propto k^4/k_{0,e}$, which is consistent with a hyperdiffusive damping. As we shall see, such separation of the electron and ion physics requires the mass ratio to be sufficiently large.  

To establish how the damping of the mass flows and magnetic fields depend on the closure parameters, we have performed one dimensional simulations initialized with sinusoidally varying, divergence-free perturbations of the respective quantities. The self-consistent exponential decay of these perturbations was used to fit flow and magnetic decay rates, $\gamma_U^-$ and $\gamma_B^-$. In particular, for the mass flow decay studies, the only non-vanishing component of the (equal) ion and electron flows are initially
$u_{\alpha z} = u_{\alpha z 0} \sin(k x)$. For the magnetic decay studies only the electrons have a non-zero initial velocity of the same form with a flow amplitude $u_{ez0}=B_{y0}k/(e\mu_0 n_e)$, corresponding to a magnetic field perturbation $B_y = B_{y0} \cos(k x)$. These simulations use the same linear box size as our dynamo simulations (namely, $L$ that defines $k_0=2\pi/L$ used for normalization), while employing ten times higher resolution for improved accuracy.  

\begin{figure}
  \centering
  \includegraphics[width=0.45\linewidth]{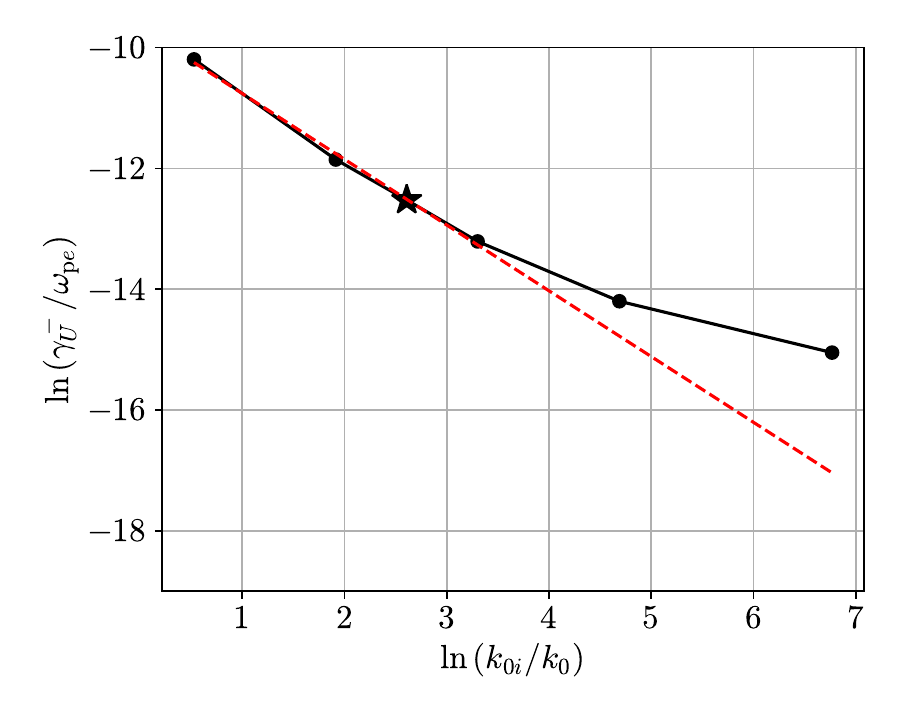}
  \put(-141,25){\includegraphics[width=0.18\linewidth, trim={5mm 5mm 5mm 5mm},clip]{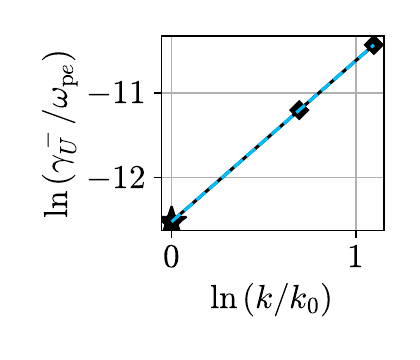}}
  \put(-25,115){a)}
  \put(-105,70){\small \textcolor{cyan}{$ k^{1.91}$}}
  \put(-55,50){\small \textcolor{red}{$ k_{0,i}^{-1.09}$}}
  \includegraphics[width=0.45\linewidth]{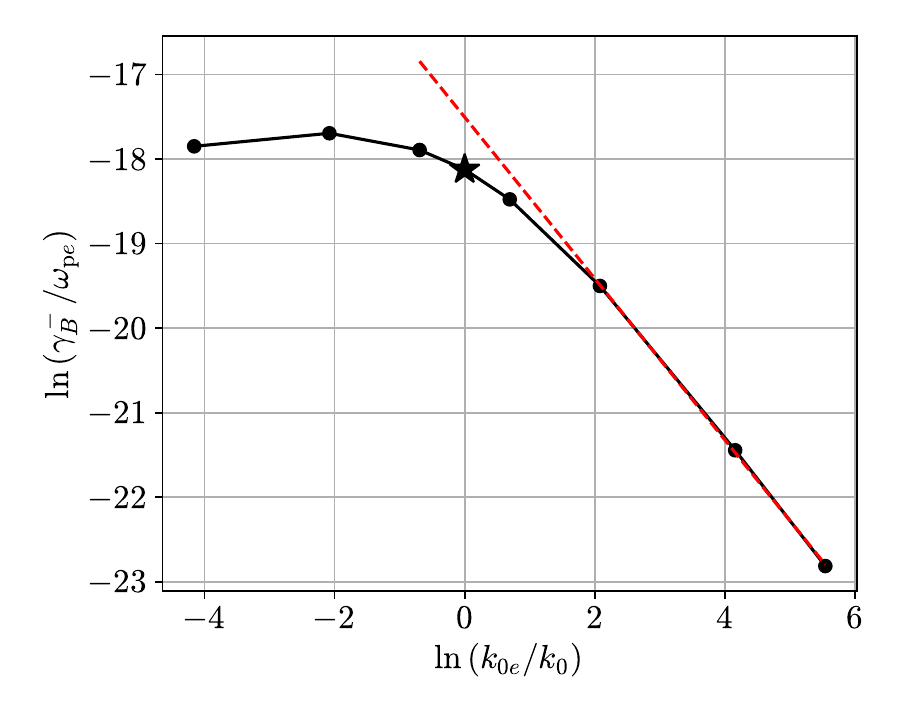}
  \put(-141,25){\includegraphics[width=0.2\linewidth, trim={1mm 5mm 5mm 5mm},clip]{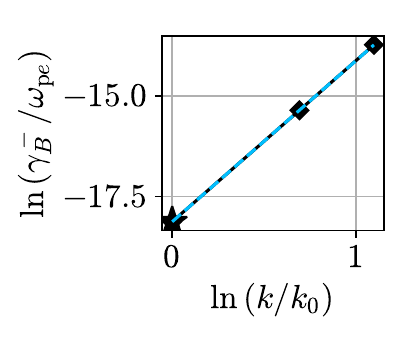}}
  \put(-25,115){b)}
  \put(-105,70){\small \textcolor{cyan}{$ k^{3.99}$}}
  \put(-70,105){\small \textcolor{red}{$ k_{0,e}^{-0.95}$}}
  \caption{\label{fig:dampings} 
  Dependence of damping rates on closure parameters (circle markers, main figure) and wavenumber (diamond markers, inset figure). In both of these scans, star markers correspond to the baseline case. Dashed lines indicate fitted relevant power-law behavior. a) Mass flow damping rate as a function of $k_{0,i}/k_0$ and $k/k_0$. b) Magnetic field damping rate as a function of $k_{0,e}/k_0$ and $k/k_0$.
  }
\end{figure}

Figure~\ref{fig:dampings} illustrates the numerically obtained dependencies of the damping rates in the vicinity of our baseline parameters (to be detailed in Sec.~\ref{sec:settings}). Over certain ranges, both $\gamma_U^-$ and $\gamma_B^-$ are approximately inversely proportional to their respective closure parameter: $\gamma_U^-\propto k_{0,i}^{-1}$ and $\gamma_B^-\propto k_{0,e}^{-1}$. For the flow damping, this dependence can be understood by a linearized analysis, with the assumptions that in the momentum equation the $m_in_i\partial_t u_{iz}$ and $\partial_x p_{ixz}$ terms nearly balance, while in the relevant $xz$ component of the pressure tensor equation, the generation of this off-diagonal pressure component by the term $p_{ixx}\partial_x u_{iz}$ is nearly compensated by its destruction by the isotropizing closure term $\propto v_{{\rm th},i}k_{0,i}p_{ixz}$. This behavior yields a dependence of the damping rate $\gamma_U^-\propto v_{{\rm th},i} k^2/k_{0,i}$, consistent with the scalings found in simulations around the baseline point. A somewhat more involved calculation of the magnetic damping rate, where Maxwell's equations are also invoked in addition to the fluid ones, yields a similar regime $\gamma_B^-\propto v_{{\rm th},e} k^4 \delta_e^2/k_{0,e}$, where $\delta_\alpha=c/\omega_{p\alpha}$ denotes the inertial length, and $\omega_{p\alpha}=\sqrt{n_\alpha q_\alpha^2/(\epsilon_0 m_\alpha)}$ the plasma frequency of species $\alpha$. We note that this differs from the exact kinetic damping of magnetic fields \citep{Mikhailovskii_1980,pusztaiPRL} that is of the form $\gamma_B^-\propto v_{{\rm th},e} k^3 \delta_e^2$; thus, our closure introduces an extra $k/k_{0,e}$ factor.

Clearly, the $1/k_{0,\alpha}$ scaling cannot hold for arbitrarily low values of the closure coefficients $k_{0,\alpha}$, because at low values, the damping rates saturate at a finite value. This behavior can be seen for $\gamma_B^-$ in Fig.~\ref{fig:dampings}b. In fact, box-scale magnetic perturbations at our baseline parameters are in the transition region to this saturation (while higher $k$ values are typically in the $\gamma_B^-\propto k_{0,e}^{-1}$ regime). However, increasing $k_{0,e}$ brings our simulations away from this saturation, providing a reduced magnetic damping, which is essential for the interpretation of the dynamo results. Concerning flow damping, we observe the damping rate to level off also at high values of $k_{0,i}$, and our baseline case is in this vicinity. It is of no concern though, as we do not perform scans in the ion closure parameter. 

\begin{figure}
  \centering
  \includegraphics[width=0.6\linewidth]{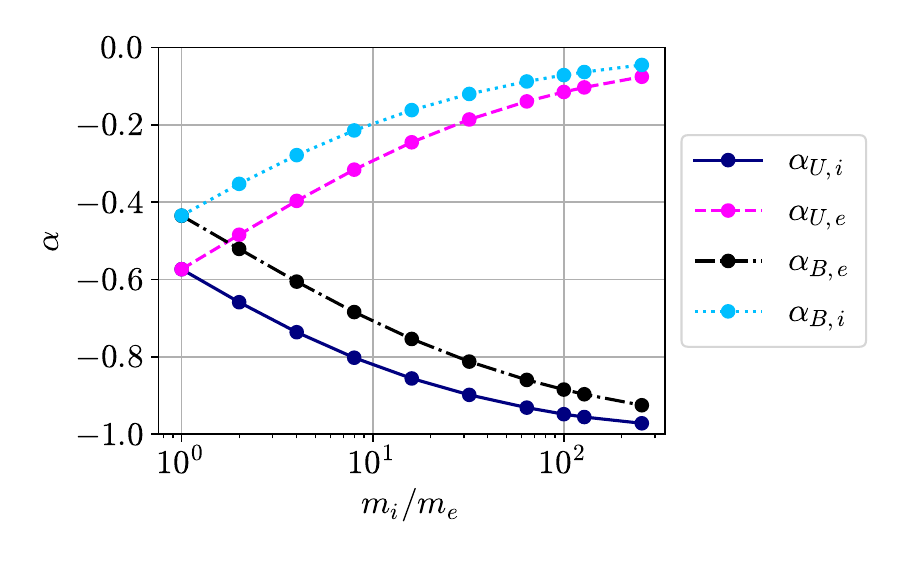}
  \caption{\label{fig:massratio} Mass-ratio dependence of the damping rate scaling exponents, defined through the power law relations $\gamma_U^-\propto k_{0,i}^{\alpha_{U,i}}k_{0,e}^{\alpha_{U,e}}$, $\gamma_B^-\propto k_{0,i}^{\alpha_{B,i}}k_{0,e}^{\alpha_{B,e}}$. }
\end{figure}

The effect of the mass ratio on the scaling of the damping rates is illustrated in Fig.~\ref{fig:massratio} in terms of the exponents defined by
\begin{equation}
\gamma_U^-\propto k_{0,i}^{\alpha_{U,i}}k_{0,e}^{\alpha_{U,e}}k^{\alpha_{U,k}}, \qquad{} \gamma_B^-\propto k_{0,i}^{\alpha_{B,i}}k_{0,e}^{\alpha_{B,e}}k^{\alpha_{B,k}}.
\end{equation}
We consider a starting point with parameters $k = k_0$ and $k_{0,i} = 13.6 k_0$, as in our baseline case, and set $k_{0,e} = 13.6 k_0$ to ensure consistency at unit mass ratio. The quantities $k_{0,i}$, $k_{0,e}$, and $k$ are then perturbed about their baseline values in magnetic-field and mass-flow damping tests to determine the above exponents. In all simulations, the electron mass is held fixed while the ion mass is varied.

The exponents characterizing the $k$ dependence remain nearly constant over the entire mass-ratio scan, with $\alpha_{U,k}\approx 2$ and $\alpha_{B,k}\approx 4$, exhibiting maximum deviations of only $1.2\%$ and $0.05\%$, respectively. As these variations are negligible, the corresponding exponents are omitted from the figure. At unit mass ratio, the effects of ion (or, in this case, positron) and electron closures on the damping of a given quantity are identical, as expected. As the mass ratio increases, $\alpha_{U,i}$ and $\alpha_{B,e}$ approach $-1$, while $\alpha_{U,e}$ and $\alpha_{B,i}$ approach $0$. In other words, at sufficiently large mass ratio, magnetic-field damping depends solely on $k_{0,e}$, whereas mass-flow damping depends solely on $k_{0,i}$, with cross-dependencies becoming negligible. This decoupling between electron and ion closures is already well converged at the mass ratio used in this study, $m_i/m_e = 100$. Finally, we note that over the mass-ratio range considered here, the sums $\alpha_{B,i} + \alpha_{U,i}$ and $\alpha_{B,e} + \alpha_{U,e}$ remain close to $-1$, with maximum deviations of $1.9\%$ and $0.8\%$, respectively.

Since the $k_{0,\alpha}\rightarrow 0$ limit of the 10-moment system is not singular, at sufficiently low values of the closure parameters $k_{0,\alpha}$ these scalings must be broken, and the damping rates saturate towards constant values. As seen in Fig.~\ref{fig:dampings}b, we are in the vicinity of such a saturation concerning $\gamma_B^-(k_{0,e})$ at our baseline parameters; however, the scan in $k_{0,e}$ we will perform moves deeper into the $\gamma_B^-\propto k_{0,e}^{-1}$ regime. Concerning the mass flows, we work in the $\gamma_U^-\propto k^2/k_{0,i}$ regime, and we keep $k_{0,i}$ fixed.

\subsection{Simulation settings}
\label{sec:settings}
To be relevant to the fast turbulent dynamo in collisionless astrophysical media, our simulations must have sustained three-dimensional flows. In reality the flows are the result of gravitational collapse in the presence of dark matter, while our simulations include forced flows of non-gravitating baryonic matter in the plasma state. To produce fast dynamo at a relatively low value of the critical magnetic Reynolds number, we drive our flows towards a simulation box-scale ($L$) Galloway-Proctor (GP) model flow \citep{Galloway_1992}. More specifically, both the electrons and ions are accelerated by the same spatio-temporally varying acceleration $\mathbf{a}_\alpha(\mathbf{x},t)=C_f \mathbf{U}_{\rm GP}(\mathbf{x},t)/\tsti$, which is proportional to the GP flow
\begin{equation}
\mathbf{U}_{\rm GP}(\mathbf{x},t) = u_0 \left[
        \begin{matrix}
        A \sin (k_0 z+\sin \omega t)+C \cos (k_0 y+\cos \omega t) \\
        A \cos (k_0 z+\sin \omega t) \\
        C \sin (k_0 y+\cos \omega t)
        \end{matrix}
        \right],
        \label{GPflow}
\end{equation} 
where $\tsti=L/\vti$ is the ion streaming time, $u_0$ is the nominal flow speed, $k_0=2\pi/L$ is the lowest wavenumber fitting the simulation box, the constants $A$ and $C$ determining the stirring geometry are chosen to be unity here, while the constant $C_f$ sets the drive strength. In our baseline case, the representative flow speed is sub-sonic $u_0=0.35 c_s$, with the sound speed $c_s=\sqrt{T_e/m_i}$, which defines the outer turnover time $\tturn=L/u_0$. As this is expected to be the characteristic timescale of dynamo growth, we use $\tturn$ for time normalization with the nominal flow speed $u_0$, while noting that the root mean square flow speed varies over time and is in general different from $u_0$. We also set the stirring frequency of the flow to $\omega=2\pi/\tturn$. 

The simulations use a mass ratio of $m_i/m_e=100$, and our baseline driving strength of $C_f=2$ and $k_{0,i}=13.6 k_0$ are chosen such that the flows remain subsonic throughout the simulation. As a reference, this value of $k_{0,i}$ corresponds to ${\rm Re}= 0.535$, while $k_{0,e}=1 k_0$ corresponds to ${\rm Rm}= 307$, according to numerical damping rate tests like those presented in Sec.~\ref{sec:reynolds}, yielding a large magnetic Prandtl number ${\rm Pm}={\rm Rm}/{\rm Re}=574$, which is relevant for the ICM. Due to numerical limitations on achievable scale separations, our approach to the ${\rm Pm}\gg 1$ limit  sacrifices the fluid cascade and modeling the system at the viscous cutoff scale. This limitation results in smooth, large-scale flows with a rapidly decaying velocity spectrum, i.e., a form of turbulence that lacks an extended inertial range due to strong damping at small scales. Despite the absence of an extended inertial range, such flows can still sustain dynamo amplification through coherent stretching \citep{Schekochihin_2004}. Moreover, it is worth emphasizing, that the dissipation in a collisionless system is not expected to be fixed: it evolves as the plasma becomes increasingly magnetized and is further modified by kinetic instabilities. 

Unlike particle-in-cell simulations, the continuum solvers of \gkyl are essentially noise-free, thus we need to prescribe magnetic perturbations from which the Weibel instability can grow. Note that due to the rapid growth of this instability, the fact that we need such a seed magnetic field is not an issue, as in astrophysical settings the instability would produce macroscopic field strengths from arbitrarily small fluctuations on timescales shorter than those of the dynamo growth. Thus, as an initial condition we prescribe a magnetic field of the form
\begin{equation}
B_{i}(t=0)=B_0 \sum_{j\ne i,n} b_{ij,n}\cos [n k_0 (x_j+L\phi_{ij,n})],
\label{magneticIC}
\end{equation}
with random relative amplitudes $b_{ij,n}$ and phases $\phi_{ij,n}$, both uniform on the unit interval, and a characteristic magnetic field strength $B_0$. Here, we only seed the 4 lowest mode numbers $n$. For our initial nominal field $B=B_0$, the thermal electron Larmor radius $\rho_e=\vte m_e/(e B)$ is $\rho_e=4.4 L$, thus the system is non-magnetized. The system is non-relativistic, since $\vte/c=0.0625$.   

A current consistent with this field, $\mathbf{j}=\mu_0^{-1}\nabla\times \mathbf{B}$, fully carried by the electrons, is also prescribed. In addition there is an initial flow velocity, equal for the electrons and the ions, given by Eq.~(\ref{GPflow}) for $t=0$. The initial ratio of the box-integrated magnetic energy and kinetic energy in the flows is $3.7\times 10^{-6}$. Our baseline box size is $L=70 \delta_i$ in each direction. With a spatial resolution of $N=140$ in each direction, such that the linear size of the finite volume is $\Delta x=L/140$, the ion inertial length is barely resolved $\delta_i=2\Delta x$, while the electron inertial length is under-resolved $\delta_e=0.2\Delta x$, along with the Debye length $\lambda_D\equiv \sqrt{\epsilon_0 T_e/(n_e e^2)}\approx 0.009\Delta x$. These scales are not a concern since the Debye scale is not playing an important role in the physics of the system, and the characteristic scales of the electron Weibel instability are significantly larger than $\delta_e$ due to the weak pressure anisotropy that develops in the system.   

We finally note that our simulations are performed in SI units with physical parameters comparable to those in \citep{pusztaiPRL}. Namely, all physical constants have their actual value, except the ion mass that is $m_i=100 m_e$; both species have initially the same scalar temperature $T_e=T_i=1\,\rm keV$, the electron density is $n_e=2.3 \times 10^{28} \,\rm m^{-3}$, along with $k_0=2.58 \times 10^5\,\rm m^{-1}$. Furthermore $B_0=1\,\rm T$. As the system has the scale invariance of a collisionless kinetic system, the fact that the plasma parameters (particularly the density) are not representative of the ICM is of no concern; we will present our results with practical normalizations.

\subsection{The electron Weibel instability in non-magnetized collisionless plasma turbulence}
\label{sec:weibel}
The Weibel instability is a non-magnetized kinetic plasma instability that arises in the presence of an anisotropic velocity distribution. Transverse perturbations in the magnetic field deflect particles in a way that produces currents reinforcing the perturbation. As a result, it tends to generate filamentary current and magnetic field structures and acts to restore pressure isotropy. It plays important roles in astrophysical and laboratory plasmas (e.g. collisionless shocks, gamma-ray bursts, laser-plasma experiments) by generating sizable magnetic fields in initially non-magnetized or weakly magnetized environments, thereby qualitatively affecting particle dynamics. The instability has both electron and ion variants and additionally arises in pair-plasmas. 

A large body of the Weibel literature is concerned with interpenetrating (often supersonic and relativistic) flow scenarios, due to its relevance for collisionless astrophysical shocks \citep{Spitkovsky_2008,Kato_2008,Medvedev_2006,Medvedev_1999}, such as supernova remnants, and laser-plasma experiments \citep{Huntington2015,Schoeffler_2014}. Pressure anisotropies appearing in such scenarios are typically substantial. However, as demonstrated by \citet{Zhou2022}, the Weibel instability also naturally occurs in non-magnetized collisionless plasma turbulence, since phase mixing in sheared flows creates pressure anisotropy $\Delta_\alpha$. For subsonic turbulence, as relevant for our topic, this anisotropy is small (see Appendix~\ref{brazil} for representative values from our simulations). 

The intuitive and conventional definition $\Delta_\alpha=p_{\|\alpha}/p_{\perp \alpha}-1$ implicitly assumes a pressure tensor with diagonal components $\{p_{\|\alpha}, p_{\perp \alpha},p_{\perp \alpha}\}$; this definition will need to be revised, since the pressure tensor in a non-magnetized system is non-gyrotropic, and has in general three distinct eigenvalues. Thus, following \citet{Zhou2022}, we use the definition $\Delta_\alpha=\sqrt{\left\langle (p_{\rm max,\alpha}/p_{\rm min, \alpha})^2\right\rangle}-1$, where $p_{\rm max,\alpha}$ is the highest eigenvalue of the pressure tensor, $p_{\rm min,\alpha}$ is the average of the two lower eigenvalues, and the angle bracket denotes the spatial (box) average. 

Since the pressure anisotropies of ions and electrons that develop in the sheared flow tend to be comparable, the \emph{electron} Weibel instability will dominate the magnetic field growth as it has higher growth rates than the ion mode. Specifically, in terms of the above anisotropy measure that is positive by construction, the fastest growing electron Weibel mode has a characteristic wavenumber $k\sim \Delta_e^{1/2}/\delta_e$ and a growth rate of $\gamma_B \sim \Delta_e^{3/2}\omega_{pe}\vte/c$. For $\Delta_e \ll 1$, the characteristic wavenumber is shifted to scales larger than the electron inertial length.


\section{Results}\label{sec:results}
\subsection{Baseline case}\label{sec:baseline}
First, we consider results for our baseline case with parameters given in Sec.~\ref{sec:settings}, to establish the qualitative features of a representative simulation, before we consider the effect of the electron closure parameter and the box size in the following subsections. 

\begin{figure}
  \centering
  \includegraphics[width=0.45\linewidth]{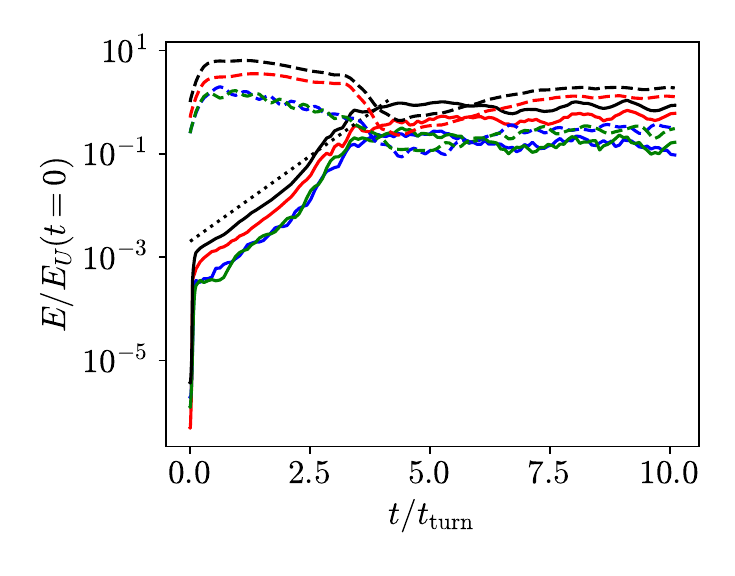}
  \put(-124,35){a)}
  \put(-102,80){$E_B$}
  \put(-86,117){$E_U$}
  \put(-132,100){\tiny $\gamma_B \tturn$}
  \put(-132,92){\tiny $= 0.76$}
  \put(-70,35){\small $k_{0,e}/k_0=1$}
  \includegraphics[width=0.48\linewidth]{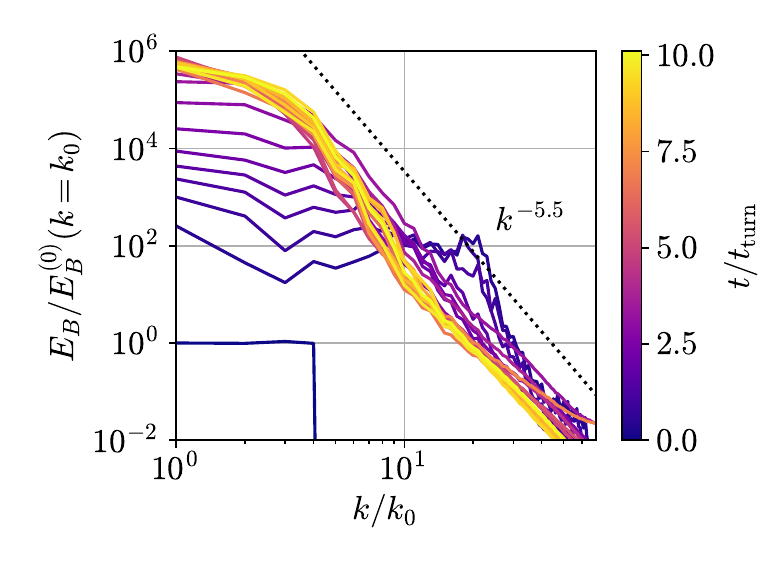}
  \put(-135,37){b)}
\caption{\label{fig:baselineLong} Magnetic field generation in the simulation with the baseline parameters. a) Time evolution of the box-integrated magnetic energy $E_B$ (solid black curve) and kinetic energy in the flows $E_U$ (dashed black), normalized to the initial value of $E_U$. With similar line styles, contributions from the $\{x, y, z\}$ components of $\mathbf{U}$ and $\mathbf{B}$ are shown in red, blue and green, respectively. The dotted line corresponds to an exponential growth of the magnetic field with a growth rate of $\gamma_B \tturn = 0.76$. b) Wave number spectra of the magnetic field energy. The wave number $k$ is normalized to $k_0$, and the spectrum is normalized to its value at $k=k_0$ for $t=0$. For reference, the dotted line indicates a power law $k^{-5.5}$. }
\end{figure}

The total magnetic field energy evolution is clearly separated into three phases, as shown in Fig.~\ref{fig:baselineLong}a (solid black line, $E_B$). The first, very short phase, $t/\tturn < 0.1$, characterized by a very rapid magnetic field growth, is dominated by the electron Weibel instability. This phase is followed by a nearly exponential growth with a growth rate comparable to the outer turnover time, approximately $\gamma_B \tturn = 0.76$, as indicated by the dotted line, which abruptly saturates around $t/\tturn = 3.4$. As the magnetic energy approaches the kinetic energy in the flow $E_U$ (dashed black line), the magnetic field brakes the flow and the flow energy drops, until eventually the two forms of energy establish near equipartition. It is noteworthy that the dynamo growth rate appears to increase towards the end of the growth phase even though the flow energy slowly decreases in time. This behavior could be caused by the increasing magnetization of the plasma leading to a reduced kinetic magnetic field damping.    

The qualitative behavior is quite similar to that observed in kinetic simulations of pair-plasma dynamos by \citet{Zhou_2024}, such as the case shown in their Fig.~1a. Besides observing the same main phases of the magnetic field evolution, we also find that, in spite of the steady forcing of the flow, the flow energy decreases after a time comparable with $\tturn$, before it finds its long time quasi-equilibrium value. However, in the results here, as a consequence of driving a flow towards the GP model flow instead of a random flow drive, the contribution of the $x$-component of the flow to the kinetic energy is larger than that of the other directions, as seen by comparing the red dashed curve ($x$-component) to the blue and green curves ($y$- and $z$-components). 

The time evolution of the magnetic energy wavenumber spectra is shown in Fig.~\ref{fig:baselineLong}b. After $t=0$, where the magnetic field is seeded only up to $k/k_0=4$, the first time point shown is already after the Weibel phase, and its high-$k$ range is dominated by a peak corresponding to magnetic energy created by the Weibel instability. The spectral peak then moves towards lower $k$, until the magnetic energy at the lowest wave number ($k=k_0$) takes over. Eventually, most of the spectrum becomes power-law-like, comparable to $k^{-5.5}$ (as indicated by the dotted line), while the lowermost $k$ range, up to around $4k_0$, exhibits a less steep spectrum, approximately $k^{-1.6}$.

\begin{figure}
  \centering
  \includegraphics[width=0.45\linewidth]{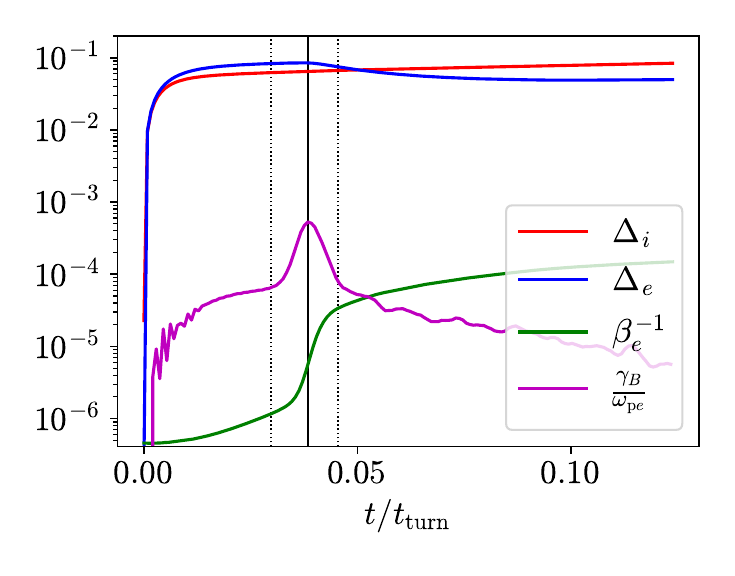}
  \put(-130,35){a)}
  \includegraphics[width=0.48\linewidth]{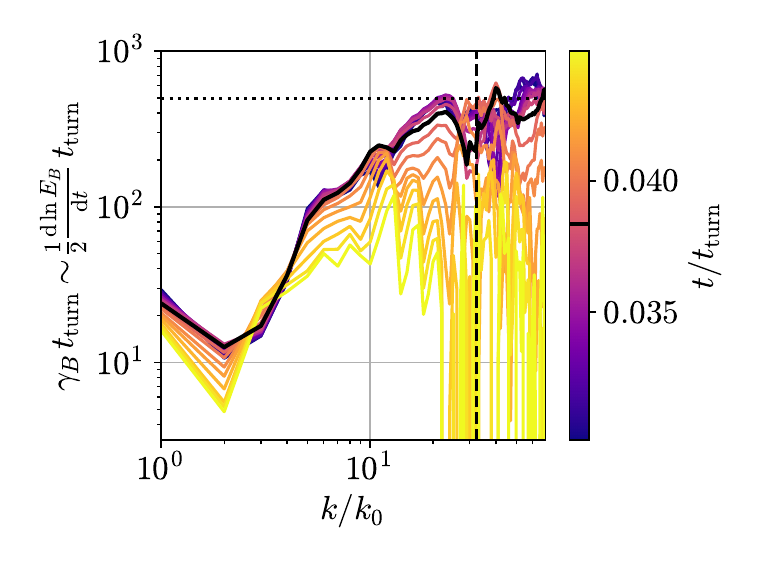}
  \put(-142,37){b)}
  \caption{\label{fig:baselineShort} Time evolution during the Weibel instability phase. a) Dimensionless quantities: Pressure anisotropy of electrons (blue, $\Delta_e$) and ions (red, $\Delta_i$), normalized magnetic field energy (green, $\beta_e^{-1}$), and instantaneous magnetic field growth rate (purple, $\gamma_B/\omega_{p,e}$). b) Instantaneous magnetic growth rate spectra ($\gamma_B\tturn$) for several time instances (color coded from blue to yellow for increasing time) across the time period of the fastest field growth, bound by the dotted vertical lines of panel a). The black curve corresponds to the time of fastest global magnetic field growth, indicated by the solid vertical line in panel a). The dotted horizontal and the dashed vertical line correspond to $\gamma_B=\Delta_e^{3/2}\omega_{pe}\vte/c$ and $k=\Delta_e^{1/2}/\delta_e$, respectively. }
\end{figure}

The dynamics of the system during the short Weibel period is illustrated in Fig.~\ref{fig:baselineShort}a, which contains similar information to Fig.~2 of \citep{Zhou2022}. The magnetic energy, now quantified via $\beta_e^{-1}$ (green curve) undergoes a very rapid growth phase approximately bounded by the dotted vertical lines, with the instantaneous growth rate of the magnetic field ($\gamma_B$, purple) peaking just before $t/\tturn=0.04$, marked by the solid line. Initially the pressure anisotropies of both species are growing in accordance with the sheared flow that is present in the system already from $t=0$. When the growth rate peaks, the instability becomes strong enough that it starts to back-react on its drive, the electron pressure anisotropy ($\Delta_e$, blue). $\Delta_e$ declines somewhat to eventually stabilize at a lower level. Meanwhile, the steady growth of the ion pressure anisotropy ($\Delta_i$, red) remains unaffected by the instability, providing one piece of evidence that the instability is indeed driven by $\Delta_e$. More details on the long-time evolution of the system in terms of electron pressure anisotropy and normalized pressure are provided in Appendix~\ref{brazil}. 

The wave number-resolved instantaneous growth rates of the magnetic field are shown for a range of time points in Fig.~\ref{fig:baselineShort}b. At the approximate time of fastest global magnetic energy growth, corresponding to the thick black line, the peak of the growth rate spectrum is located in the vicinity of $k=\Delta_e^{1/2}/\delta_e$ and takes a value comparable to $\gamma_B=\Delta_e^{3/2}\omega_{pe}\vte/c$, indicated by the dashed vertical and the dotted horizontal lines respectively. These are the theoretical scalings from \citep{Zhou_2024}, without their specific order unity prefactors. The purpose of these is just to illustrate that the magnetic field growth is consistent with being produced by the electron Weibel instability. An accurate numerical comparison would require theory for a system similar to our simulated scenario, but even then, differences due to our simulation being fluid are expected.      

\begin{figure}
  \centering
  \includegraphics[width=1.0\linewidth]{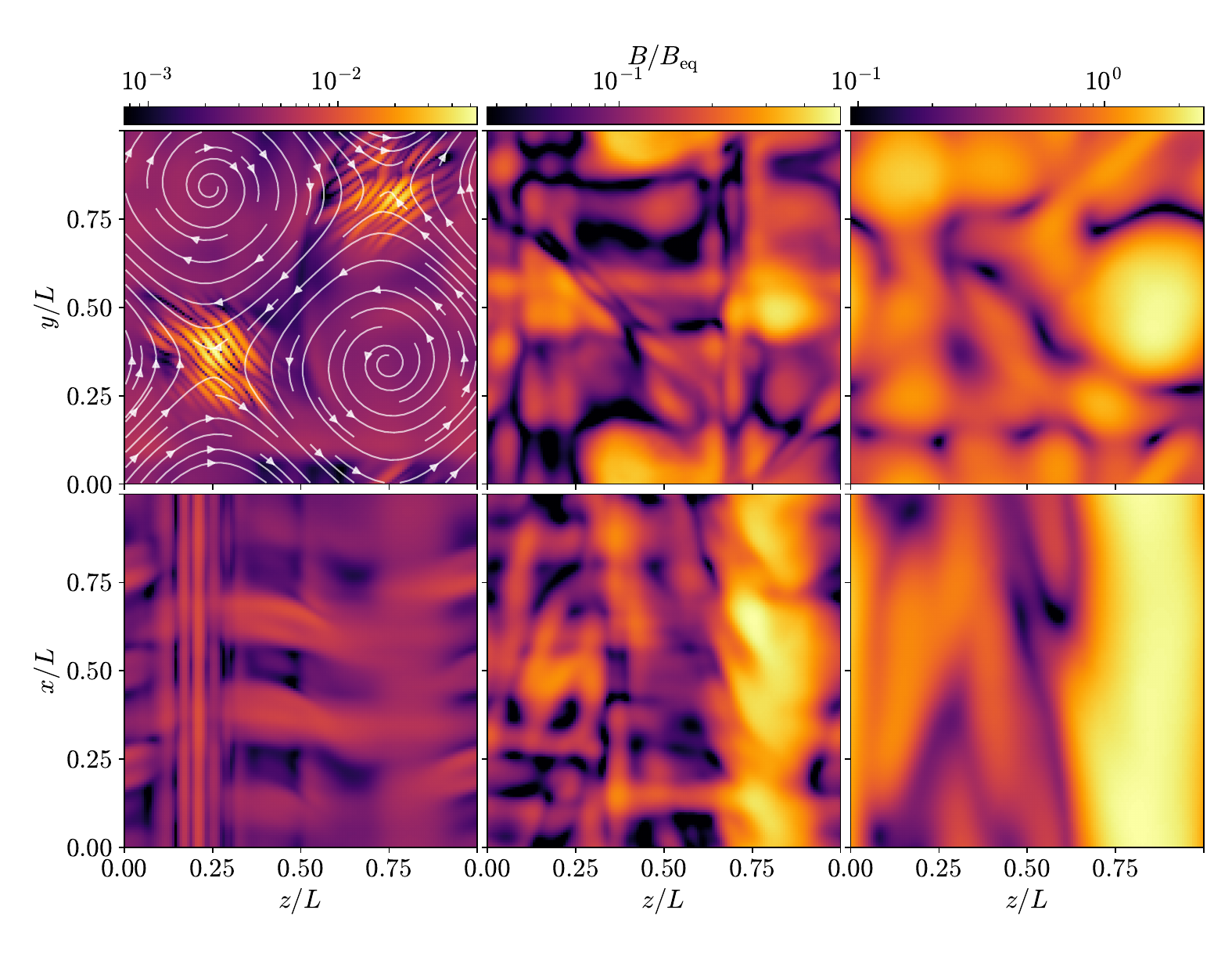}
  \put(-340,245){\textcolor{white}{\large a)}}
  \put(-227,245){\textcolor{white}{\large b)}}
  \put(-114,245){\textcolor{white}{\large c)}}
  \put(-340,130){\textcolor{white}{\large d)}}
  \put(-227,130){\textcolor{white}{\large e)}}
  \put(-114,130){\textcolor{white}{\large f)}}
  \caption{\label{fig:cutsL} Magnitude of the magnetic field in 2-D cuts of the simulation domain, taken a) and d) at the time of the fastest magnetic growth during the Weibel phase, $t/\tturn=0.038$, b) and e) in the middle of the dynamo growth phase, $t/\tturn=2.0$, and c) and f) in the saturated phase, $t/\tturn=10.0$. a)-c) are cuts at $x=L/2$, while d)-f) are cuts at $y=L/2$ (the latter are morphologically similar to constant $z$ cuts, not shown). The normalizing, ``equipartition'' magnetic field $B_{\rm eq}$ is defined such that $B_{\rm eq}^2/(2\mu_0)$ equals the box-averaged kinetic energy density in the bulk flows at $t=0$.}
\end{figure}

The morphology of the magnetic field strength is shown in Fig.~\ref{fig:cutsL} for the three phases of the magnetic field evolution. The left panels, taken at the time of fastest magnetic growth in the Weibel phase, show characteristic Weibel filamentation coincident with the regions of strongest shear in the GP flow structure (stream plotted with white arrows). Recall from Eq.~(\ref{GPflow}) that the flow is independent of $x$, and it forms two circulations in the $y$-$z$ plane around the elliptic fixed points located at $\{z,y\}\approx\{0.25,0.84\}$ and $\{z,y\}\approx\{0.75,0.34\}$. In these nearly circular regions, the magnetic field remains relatively low, with some structure around the boundary of the circulations, as shown in Fig.~\ref{fig:cutsL}a. The strongest magnetic energy growth is instead observed in the vicinity of the hyperbolic fixed points located around $\{z,y\}\approx\{0.25,0.34\}$ and $\{z,y\}\approx\{0.75,0.84\}$. It is in fact in these regions where the shear strain of the flow is strongest\footnote{Quantitatively, these hyperbolic fixed-point areas produce the highest negative values of the Q-parameter, also referred to as the second invariant of the flow gradient tensor \citep{Chong1990}.}. 
In this $z$-$y$ plane, the magnetic filaments are aligned parallel to the unstable separatrix. In the $z$-$x$ plane, Fig.~\ref{fig:cutsL}d, the same filaments, which form planar structures in three dimensions, are elongated along the $x$ direction, best seen around $z=0.25$, which is close to one of the highly strained regions. In the rest of the plane, the striations are elongated mostly in the $z$ direction; they have longer wavelengths and are caused by the sheared flows in the $x$-direction. 

In the dynamo growth phase, shown in Fig.~\ref{fig:cutsL}b and e, the localized fine scale striations corresponding to the electron Weibel instability disappear, and no clear directionality is left in the $z$-$y$ plane. The characteristic scale of the fluctuations is clearly larger than that of the Weibel striations. Due to the special form of the flow forcing, the fluctuations are not completely isotropic even in the $z$-$y$ plane, rather some structure appears to be aligned with the coordinate axes. This preferential alignment in the $z$-$y$ plane is not apparent in the saturated state, shown in Fig.~\ref{fig:cutsL}c and f. However, there is an alignment with the special direction $x$, which, being a symmetry direction of the forcing, consistently supports the strongest flows (as also seen by comparing the corresponding flow energies: the red dashed curve to the blue and green ones in Fig.~\ref{fig:baselineLong}a).  

\subsection{Increased electron pressure isotropization} \label{sec:k0escan}

\begin{figure}
  \centering
  \includegraphics[width=0.45\linewidth,trim={0 0.8cm 0 0.5cm},clip]{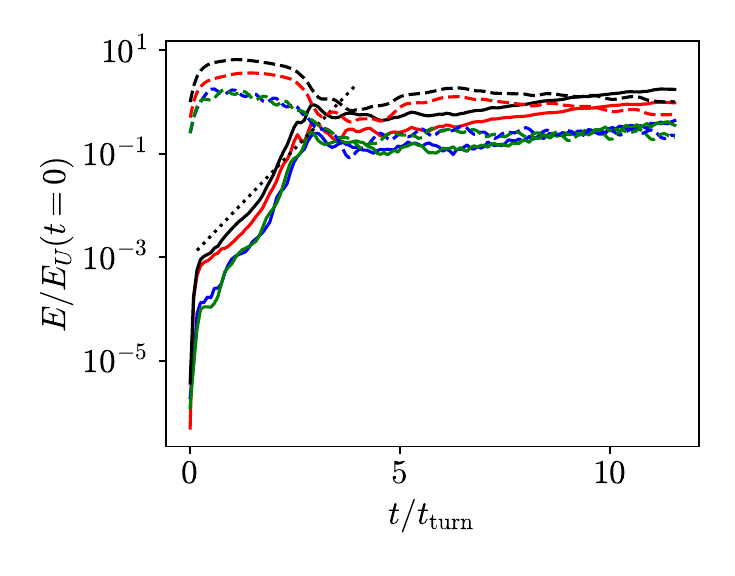}
  \put(-124,25){a)}
  \put(-105,70){$E_B$}
  \put(-97,107){$E_U$}
  \put(-133,90){\tiny $\gamma_B \tturn$}
  \put(-133,82){\tiny $=0.97$}
  \put(-70,25){\small $k_{0,e}/k_0=2$}
  \includegraphics[width=0.48\linewidth,trim={0 0.8cm 0 0.5cm},clip]{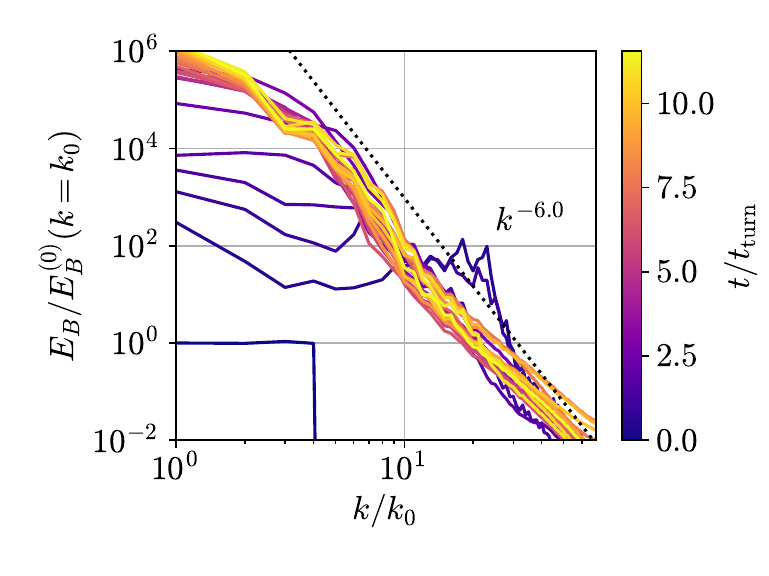}
  \put(-135,25){b)} \\
  \includegraphics[width=0.45\linewidth,trim={0 0.8cm 0 0.5cm},clip]{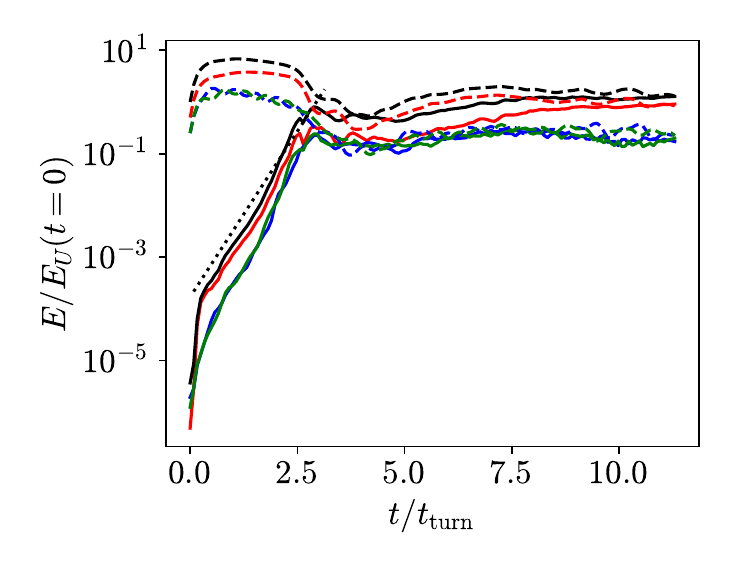}
  \put(-124,25){c)}
  \put(-105,70){$E_B$}
  \put(-97,107){$E_U$}
  \put(-133,90){\tiny $\gamma_B \tturn$}
  \put(-133,82){\tiny $=1.47$}
  \put(-70,25){\small $k_{0,e}/k_0=4$}
  \includegraphics[width=0.48\linewidth,trim={0 0.8cm 0 0.5cm},clip]{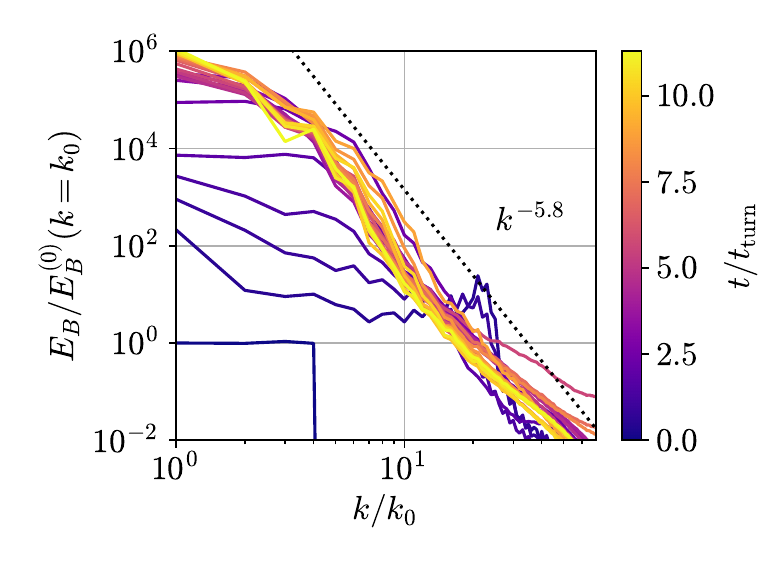}
  \put(-135,25){d)} \\
  \includegraphics[width=0.45\linewidth,trim={0 0.8cm 0 0.5cm},clip]{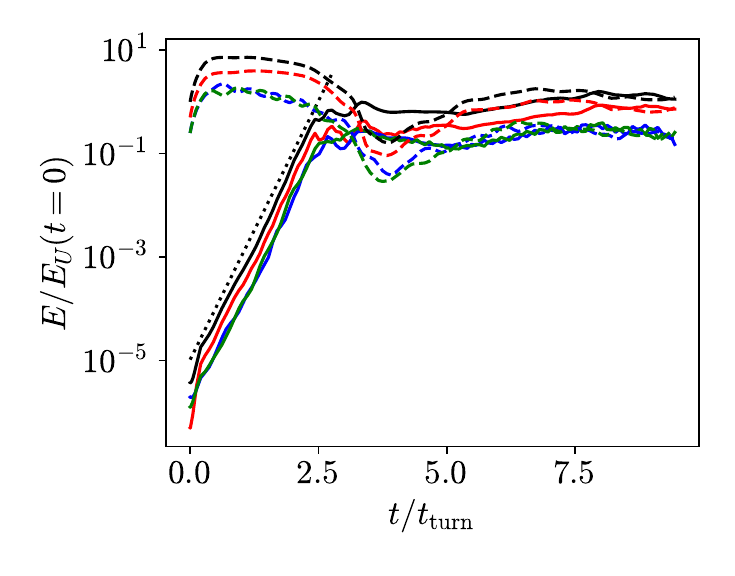}
  \put(-124,25){e)}
  \put(-105,70){$E_B$}
  \put(-92,107){$E_U$}
  \put(-132,88){\tiny $\gamma_B \tturn$}
  \put(-132,80){\tiny $= 2.3$}
  \put(-70,25){\small $k_{0,e}/k_0=8$}
  \includegraphics[width=0.48\linewidth,trim={0 0.8cm 0 0.5cm},clip]{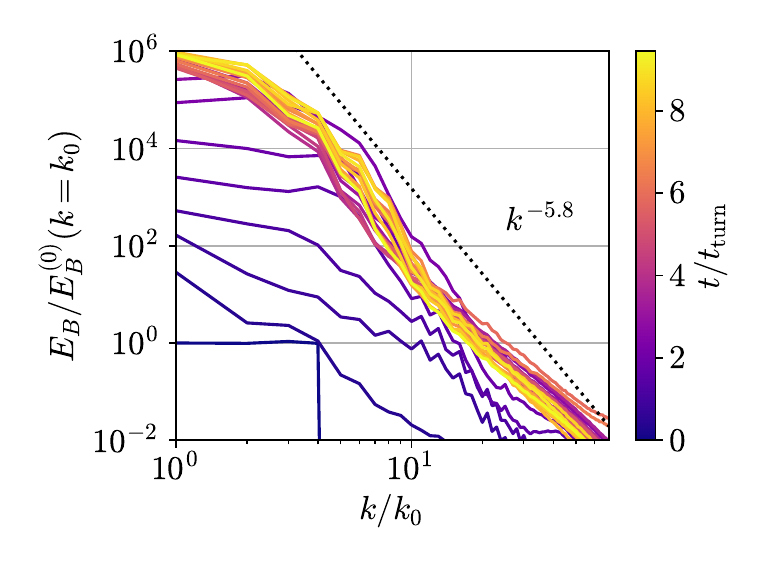}
  \put(-135,25){f)} \\
  \includegraphics[width=0.45\linewidth,trim={0 0.8cm 0 0.5cm},clip]{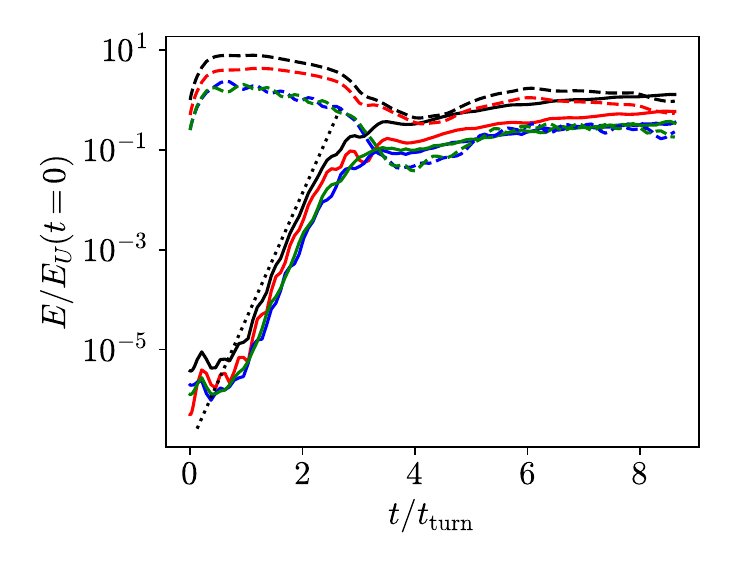}
  \put(-124,25){g)}
  \put(-97,70){$E_B$}
  \put(-87,107){$E_U$}
  \put(-132,80){\tiny $\gamma_B \tturn$}
  \put(-132,72){\tiny $= 2.9$}
  \put(-70,25){\small $k_{0,e}/k_0=32$}
  \includegraphics[width=0.48\linewidth,trim={0 0.8cm 0 0.5cm},clip]{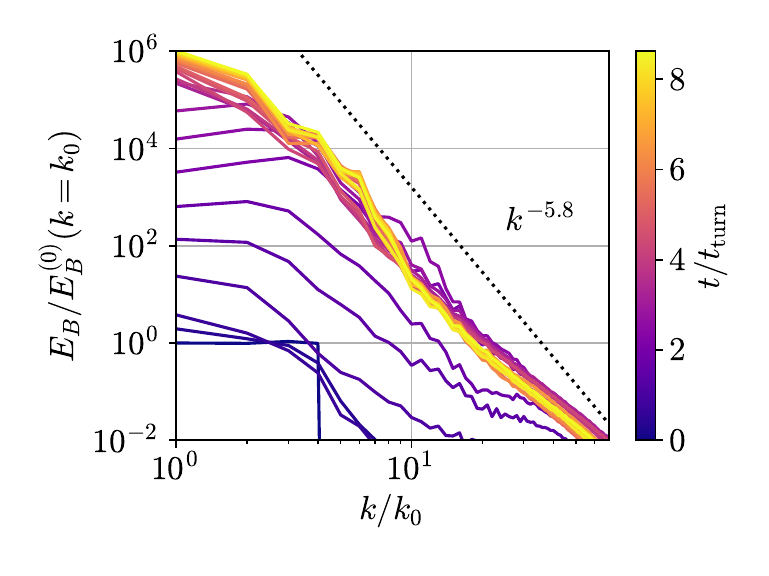}
  \put(-135,25){h)}
\caption{\label{fig:k0escan} Magnetic field generation in simulations using $k_{0,e}/k_0=\{2,\,4,\,8,\,32\}$, shown in different rows. Left column: Time evolution of the box-integrated magnetic energy $E_B$ (solid black curve) and kinetic energy in the flows $E_U$ (dashed black), normalized to the initial value of $E_U$. With similar line styles, contributions from the $\{x, y, z\}$ components of $\mathbf{U}$ and $\mathbf{B}$ are shown in red, blue and green, respectively. Exponential magnetic growth is indicated by dotted lines with growth rates provided in the figure. Right column: Wave number spectra of the magnetic field energy. The wave number $k$ is normalized to $k_0$ and the spectra are normalized to its value at $k=k_0$ for $t=0$. Dotted lines indicate power laws for reference. }
\end{figure}

We will now consider the effect of increasing electron pressure isotropization from its baseline value $k_{0,e}/k_0=1$. Information similar to that in Fig.~\ref{fig:baselineLong}, but for $k_{0,e}/k_0=\{2,\,4,\,8,\,32\}$, is shown in Fig.~\ref{fig:k0escan}. As the strength of the isotropization is increased, and thus the drive of the Weibel instability is reduced, we observe a diminishing Weibel growth phase. Up to the $k_{0,e}/k_0=4$ case, we still observe a magnetic growth rate that increases in time during the otherwise nearly exponential dynamo growth phase, but it is not present in the higher $k_{0,e}$ cases. Perhaps more importantly, the representative growth rate increases with increasing $k_{0,e}$, as indicated by the dotted lines and $\gamma_B\tturn$ values in the figure. This behavior is consistent with the magnetic damping decreasing as $1/k_{0,e}$. To draw an analogy with the dynamo in resistive magnetohydrodynamics, we may think of the reduced magnetic damping as an increased magnetic Reynolds number. A fast dynamo produces an increasing, and eventually saturated growth rate with increasing magnetic Reynolds number, which is qualitatively consistent with the increasing $\gamma_B$ we observe for increasing $k_{0,e}$.

Similar to the $k_{0,e}/k_0=1$ case, when the magnetic field energy reaches dynamically significant levels, it back-reacts on the flow, leading to a temporary reduction of the flow energy. Eventually, the flow energy grows back to nearly its initial value, and the two forms of energy reach near equipartition. Again, due to the type of flow we drive, the $x$ component of both the flow and the magnetic energy provide the largest contribution to the respective energies (red lines).  

The diminishing Weibel phase with increasing $k_{0,e}$ is also reflected in the magnetic energy spectra, where the pronounced Weibel peak seen for $k_{0,e}/k_0=1$ is not clearly present anymore. The magnetic spectra develop similarly strong dependence on $k$ as the $k_{0,e}/k_0=1$ case, and no clear trend is observed in terms of the spectral shape as $k_{0,e}$ is increased. The spectra appear to exhibit two changes in slope (most clearly seen in the $k_{0,e}/k_0=8$ case), with the intermediate range displaying the steepest spectral index. However, given the limited spatial-scale separation and available statistics, this feature should be interpreted with caution.

As $k_{0,e}$ is increased, the growth rate and characteristic wavenumber of the electron Weibel instability is reduced, as shown in Appendix~\ref{app:weibel}. At some point the spatio-temporal scales of the electron Weibel instability would become comparable to that of the ion Weibel instability, which is a scenario not anticipated based on current physical understanding. 

Finally, we note that our simulations capture the evolution from electron Weibel field generation to a saturated dynamo. The absence of an otherwise expected ion Weibel phase is primarily due to the limited domain size, which does not allow for sufficiently large and coherent regions of flow and ion pressure anisotropy required to trigger the ion-scale instability. We also recall that the characteristic scale of the ion instability depends on the magnitude of the ion pressure anisotropy, which in turn is affected by the relatively strong ion closure used in our simulations. We therefore interpret the absence of an ion Weibel phase mainly as a limitation of scale separation.

\section{Summary and discussion}\label{sec:conclusions}
Recent progress in understanding magnetic-field generation in large, weakly collisional astrophysical systems has demonstrated the need to capture physics beyond magnetohydrodynamics. At the same time, fully kinetic investigations remain severely constrained in both parameter coverage and dynamic range due to the extreme numerical demands of this inherently three-dimensional, multiscale problem. Here, we leverage the capabilities of the 10-moment collisionless fluid solver of \gkyl to provide an alternative approach to the problem, with physics fidelity and computational complexity between the mentioned extremes. The model allows us to perform simulations, at modest numerical expense, that bridge the spatiotemporal scales between seed generation by the electron Weibel instability and a subsequent, saturating dynamo process, at a mass ratio well above unity and a non-relativistic temperature. 

The results are qualitatively similar to previous fully kinetic simulations performed in a pair plasma. The process starts with a distinct phase dominated by Weibel magnetic field growth, destabilized by electron pressure anisotropy that, in turn, is generated by phase mixing of sheared flows. The phase is followed by dynamo growth of the magnetic energy on outer-turnover timescales, saturating at dynamically significant levels, in equipartition with the kinetic energy in the bulk flow. However, unlike pair-plasmas, the role of the two particle species is clearly separated: The first sign of this is that the electron Weibel instability only regulates the electron pressure anisotropy upon saturation, while it leaves the ion anisotropy unaffected. More generally, as shown in Sec.~\ref{sec:reynolds}, the scale-dependent damping of the mass flow is determined by the ion heat flux as the mass flow is dominantly carried by the ions, while the magnetic field damping is governed by the electron heat flux as the currents are mostly carried by electrons.

This study focuses on the early, electron-driven phase of magnetogenesis, and the simulation design is not suited to capture a subsequent ion Weibel phase, which is nevertheless expected to occur. Exploring the self-consistent evolution of both electron and ion Weibel instabilities in such shear-flow-induced anisotropy scenarios is an important but computationally demanding task left for future work. Importantly, these results should not be interpreted as implying that magnetogenesis is controlled solely by electron-scale physics; ion-scale processes are expected to play an essential role.

In the 10-moment model, the heat fluxes are not self-consistent, rather they are provided by a closure model. In particular, we use the pressure isotropization closure, with one free parameter per species, $k_{0,\alpha}$. These free parameters then control the damping of flows and magnetic fields. The flow damping caused by this heat flux model is found to be diffusive ($\propto k^2$), in analogy with viscous damping, and the magnetic field damping is superdiffusive ($\propto k^4$). By adjusting $k_{0,e}$ and $k_{0,i}$ we can explore various effective magnetic and fluid Reynolds numbers. We target ${\rm Pm}\gg 1$, to be relevant to intracluster media, using ${\rm Re}<1$ and ${\rm Rm}\gg 1$.

Increasing the electron isotropization strength $k_{0,e}$ and thereby the effective ${\rm Rm}$ value (as the magnetic field damping is inversely proportional to $k_{0,e}$), we observe increasing dynamo growth rates, similar to fast dynamos in resistive MHD. In addition, the simulations become more MHD-like as $k_{0,e}$ is increased, in the sense that the pressure remains nearly isotropic, and as a consequence, the Weibel phase diminishes in length.

Adopting the 10-moment modeling approach opens up several avenues for further investigation. Here, we started to explore the effect of the free parameters of the heat flux closures; however, these may instead be chosen such as to optimally account for some aspects of the physics, such as accurately capturing the wave number spectrum or the saturation level of the Weibel instability. Improving the heat flux closure model through e.g., theoretical \citep{HammettPerkins} or data driven approaches \citep{Ingelsten_2025,Huang_2025,Barbour_2025} would allow a more accurate representation of some physical process deemed to be of central importance. It is worth noting though, that magnetogenesis provides a particularly challenging setting for closure construction. The applicability of any fixed closure is inherently limited, as the plasma undergoes a transition from unmagnetized to magnetized states, during which kinetic effects (e.g., phase mixing and heat transport) change qualitatively. Constructing closures that remain accurate across these regimes is a challenging open problem. 

The reduced computational cost compared to the kinetic approach enables a dedicated study of mass-ratio effects, providing an essential input for extrapolating these results to the physical mass ratio of an electron-proton plasma. The accuracy and the possibilities to capture various kinetic effects can be enhanced by representing various particle populations (e.g., hot and cold) as separate species, with only a limited increase in computational complexity. The larger numerical flexibility is also beneficial for improving the realism of the turbulence driver, by e.g., modeling gravitational collapse with a dark matter component that is only gravitationally coupled to the plasma.

\section*{Acknowledgments} 
The authors are grateful to Muni Zhou for a fruitful discussion, as well as Sarah Newton and T\"{u}nde F\"{u}l\"{o}p for constructive comments on the manuscript.
The computations were enabled by resources provided by the National Academic Infrastructure for Supercomputing in Sweden (NAISS), partially funded by the Swedish Research Council through grant agreement No.~2022-06725. The authors also acknowledge the Texas Advanced Computing Center (TACC) at The University of Texas at Austin for providing computational resources that have contributed to the research results reported within this paper.

\section*{Funding} 
The work was supported by the Swedish Research Council (Dnr.~2021-03943) and the Knut and Alice Wallenberg foundation. 
J.~Juno, A.~Hakim, J.~M.~TenBarge, and the development of \gkyl were partly funded by the NSF-CSSI program, Award No. 2209471. J.~Juno and A.~Hakim were also supported by the U.S. Department of Energy under Contract No. DE-AC02-09CH1146 via LDRD grants. J.~M.~TenBarge was also supported by NASA grant 80NSSC23K0099.

\section*{Declaration of Interests}
Competing interests: The authors declare none.

\appendix

\section{Weibel instability in the 10-moment system}
\label{app:weibel}
Here we provide a reference point for how the Weibel instability is captured in the 10-moment model implemented in \gkyl. Previous work by \citet{Kuldinow_2025} demonstrated that the 10-moment system retains the Weibel instability. In that study, a zero heat flux closure was employed, for which the resulting dispersion relation is
\begin{equation}
    1-\frac{k^2c^2}{\omega^2}-\sum_\alpha \frac{\omega_{p\alpha}^2}{\omega^2}\left[1+\frac{T_{\alpha\perp}}{2T_{\alpha\|}}\frac{1}{\zeta_\alpha^2-1/2}\right]=0. 
\end{equation}
This expression, consistent with \citep{Sarrat_2016}, assumes bi-Maxwellian distributions for each species $\alpha$, where $T_{\alpha\|}$ and $T_{\alpha\perp}$ denote the temperatures parallel and perpendicular to the wavevector $k$, respectively. Here $\zeta_\alpha = \omega/(k v_{{\rm th},\alpha})$. 

For comparison, the corresponding kinetic dispersion relation is
\begin{equation}
    1-\frac{k^2c^2}{\omega^2}-\sum_\alpha \frac{\omega_{p\alpha}^2}{\omega^2}\left[1+\frac{T_{\alpha\perp}}{2T_{\alpha\|}}Z'(\zeta_\alpha)\right]=0, 
\end{equation}
where $Z'(\zeta_\alpha) = -2\left[1+\zeta_\alpha Z(\zeta_\alpha)\right]$ is the derivative of the plasma dispersion function $Z(\zeta_\alpha)$. The difference between the two expressions scales as $\sim \zeta_\alpha^{-4}$, so the 10-moment result is expected to be accurate in the limit of large $|\zeta_\alpha|$, i.e., for sufficiently large growth rates and long wavelengths. While the zero heat flux closure accurately captures the range of unstable wavenumbers, it overestimates the linear growth rate, particularly when the instability is weakly driven (i.e., when the initial pressure anisotropy is small). The nonlinear saturation amplitude, however, is reproduced reasonably well.

In our current context, where the pressure anisotropy is generated by sub-sonic sheared flows, the anisotropy remains small, therefore the zero heat flux closure is expected to be inaccurate. Motivated by this, we explore whether the heat flux closure parameter can be tuned to better reproduce selected properties of the instability.

To facilitate comparison with previous work, we adopt the setup used in the Supplementary Information of \citep{Zhou2022}. Specifically, we consider a pair-plasma subject to a box-scale acceleration of the form $a_y = a_0 \sin(k_0 x)$ in a two-dimensional simulation. The acceleration amplitude is parameterized as $a_0 = S_0 \pi^2 \theta_e c^2 / L$, where $\theta_e = T_e / (m_e c^2)$. We use their baseline parameters $\theta_e = 1/16$, $S_0 = 0.2$, and $L/d_e = 128$. 

Because our simulations lack particle noise, we initialize the magnetic field using Eq.~(\ref{magneticIC}) (excluding terms that depend on $z$), seeding the 128 lowest wavenumbers with an amplitude $B_0$ such that $\beta_e^{-1} = 1.41 \cdot 10^{-6}$ initially. The closure parameter is taken to be identical for both species, so we report only the electron value $k_{0,e}$.

\begin{figure}
    \includegraphics[width=0.325\linewidth, trim={5mm 5mm 5mm 5mm}, clip]{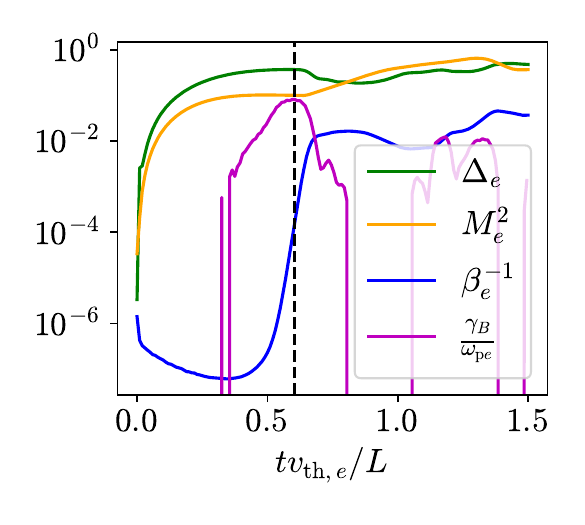}
    \put(-103,98){\small (a)}
    \includegraphics[width=0.325\linewidth, trim={5mm 5mm 5mm 5mm}, clip]{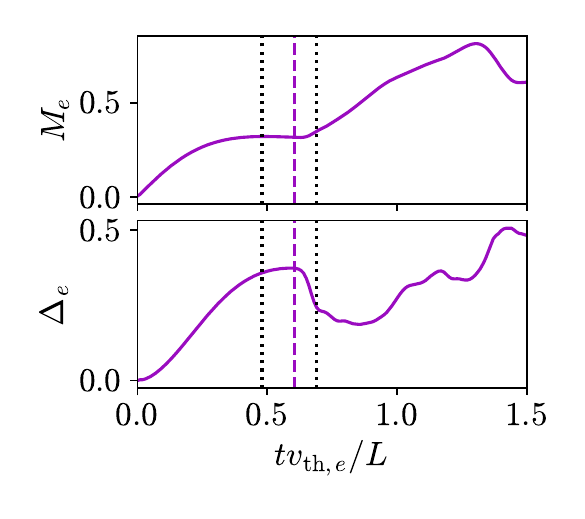}
    \put(-98,98){\small (b)}
    \includegraphics[width=0.35\linewidth, trim={5mm 5mm 5mm 5mm}, clip]{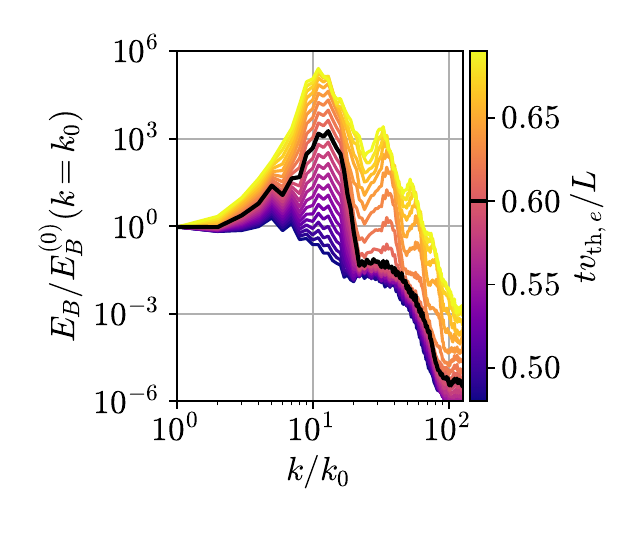}
    \put(-98,98){\small (c)}
    \caption{\label{fig:Z22_2D_mr2_k0e_1p3} a) Time evolution of pressure anisotropy, $\Delta_e$ (green), squared Mach number $M_e^2=\langle u_e^2 / v_{{\rm th},e}^2 \rangle$ (orange), normalized magnetic energy $\beta_e^{-1}$ (blue), and instantaneous growth rate $\gamma_B/\omega_{pe}$ (magenta), shown on a logarithmic scale. b) $M_e$ and $\Delta_e$ on a linear scale. c) Time evolution of the wavenumber spectrum. The time of fastest magnetic energy growth is indicated by the vertical dashed line in panel a), corresponding to the black curve in panel c).} 
\end{figure}

Figure~\ref{fig:Z22_2D_mr2_k0e_1p3} shows the evolution of plasma parameters and the magnetic field spectrum for a representative case with $k_{0,e}/k_0 = 1.3$. Note that, in this appendix only, we adopt the definition $v_{{\rm th},e} = \sqrt{T_e / m_e}$ to facilitate comparison with \citep{Zhou2022}. The time evolution of $M_e$, $\Delta_e$, and $\gamma_B$ is in good qualitative agreement with Fig.~S3 of \citep{Zhou2022}. The initial decay of $\beta_e$ observed in our simulations is a consequence of the absence of a particle noise floor. 

The time evolution of the wave number spectrum of the magnetic field, shown in Fig.~\ref{fig:Z22_2D_mr2_k0e_1p3}c shows a peculiar behavior: before the growth rate reaches its maximum (corresponding to the black curve), wave growth is dominated by the most unstable wavenumber, slightly above $k/k_0=10$. Afterwards, when the saturation starts, additional distinct peaks at higher $k$ start to grow, significantly faster than the growth of the most unstable mode (compare the spacing of the curves). These are odd harmonics of the most unstable mode and reflect that the magnetic field perturbation begins to approach a more square-wave like from, separated by narrow current sheets. This is consistent with the kinetic saturation behavior studied by \citet{Martyanov2008}.     

We characterize the instability using three quantities: (i) the maximum instantaneous growth rate, $\gamma_{B,{\rm max}}$, measured at the time indicated by the dashed vertical line in Fig.~\ref{fig:Z22_2D_mr2_k0e_1p3}a; (ii) the characteristic wavenumber, $k_{E_B,{\rm max}}$, defined as the wavenumber at which the magnetic energy spectrum peaks at the same time (i.e., the peak of the black curve in Fig.~\ref{fig:Z22_2D_mr2_k0e_1p3}c); and (iii) the maximum pressure anisotropy, $\Delta_{e,{\rm max}}$, attained during the growth phase (between the dotted lines in Fig.~\ref{fig:Z22_2D_mr2_k0e_1p3}b). All three quantities decrease as the closure parameter $k_{0,e}$ is increased.

\begin{figure}
  \centering
  \includegraphics[width=0.45\linewidth]{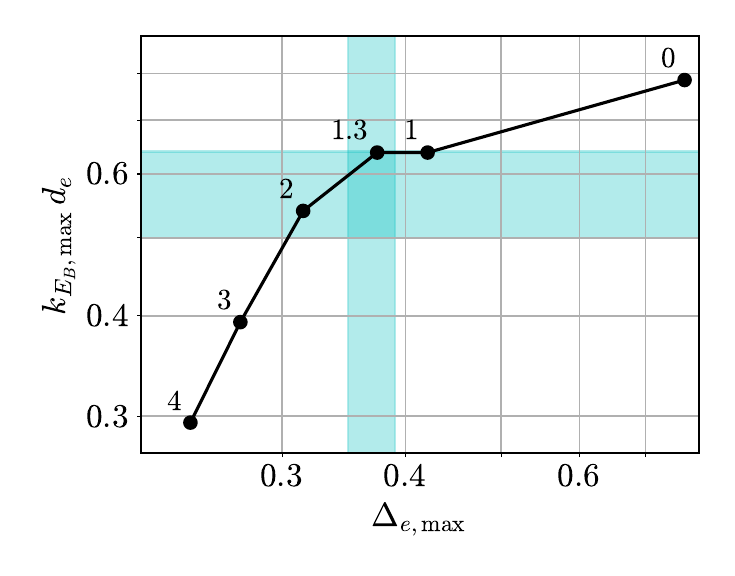}
  \includegraphics[width=0.45\linewidth]{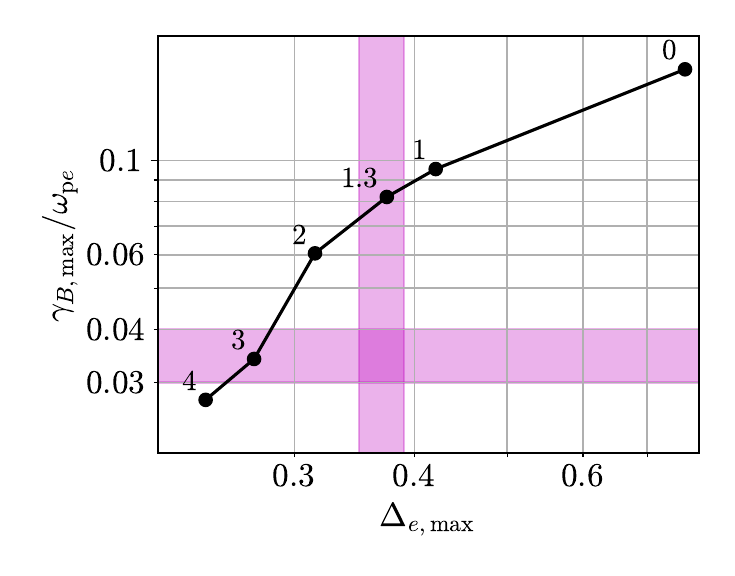}
  \caption{\label{fig:matching} 
Dependence of key Weibel instability characteristics on the heat flux closure parameter. Marker labels indicate values of $k_{0,e}/k_0$ ranging from $0$ to $4$. Shaded regions show estimates from the kinetic results presented in supplementary Fig.~S5 of \citep{Zhou2022}. (a) $\Delta_{e,{\rm max}}$ and $k_{E_B,{\rm max}}$. (b) $\Delta_{e,{\rm max}}$ and $\gamma_{B,{\rm max}}$.} 
\end{figure}

Figure~\ref{fig:matching} shows how these characteristics vary as $k_{0,e}/k_0$ is increased from $0$ to $4$. We find that $\Delta_{e,{\rm max}}$ and $k_{E_B,{\rm max}}$ can be simultaneously matched to the kinetic estimates (green shaded regions) for $k_{0,e}/k_0 \approx 1.3$, which is the value used in Fig.~\ref{fig:Z22_2D_mr2_k0e_1p3}. Matching the growth rate instead requires a larger value, $k_{0,e}/k_0 \approx 3$. 

In the context of magnetogenesis, the former two quantities are arguably more relevant, as they determine the pressure anisotropy and the characteristic magnetic coherence scale, which together provide initial conditions for the dynamo. By contrast, the Weibel growth rate is typically much faster than dynamo timescales, so even a factor $\sim 2$ discrepancy is unlikely to significantly affect the system's evolution. We therefore conclude that choosing moderately small values of $k_{0,e}$ allows the 10-moment model to qualitatively reproduce key features of the kinetic system.

\section{Evolution of pressure anisotropy and normalized pressure}
\label{brazil}

\begin{figure}
  \centering
  \includegraphics[width=1.0\linewidth]{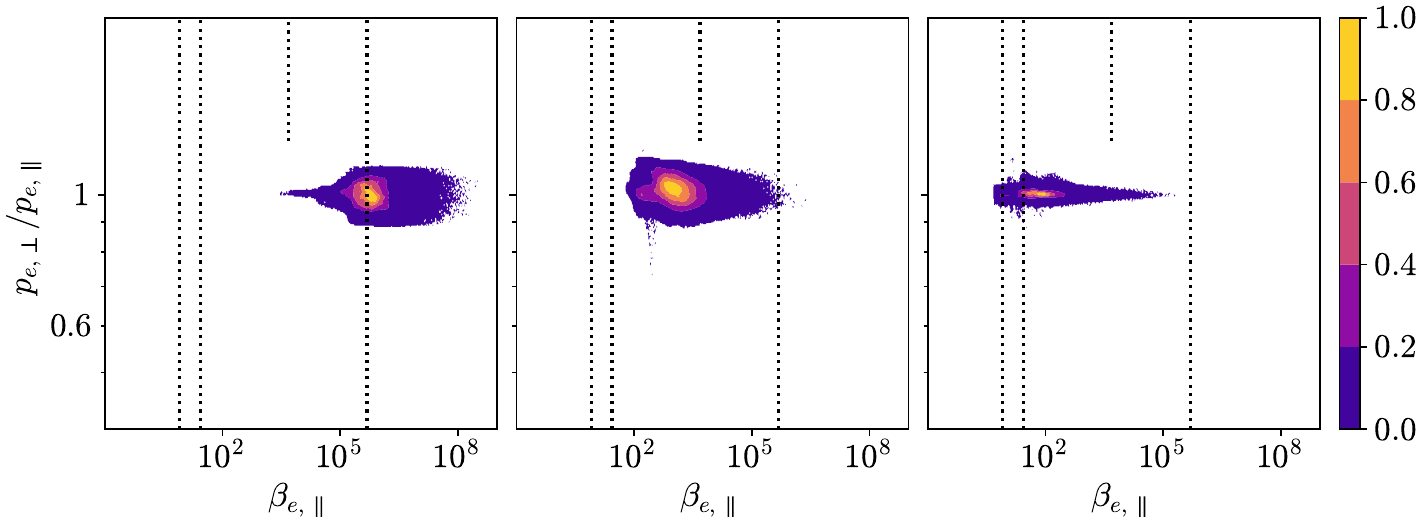}
  \put(-352,30){a)}
  \put(-241,30){b)}
  \put(-130,30){c)}
  \put(-284,30){\tiny \rotatebox{90}{$\rho_e=L$}}
  \put(-313,110){\tiny \rotatebox{90}{$\rho_i=L$}}
  \put(-328,30){\tiny \rotatebox{90}{$\rho_e=k_{\rm Weibel}^{-1}$}}
  \put(-341,30){\tiny \rotatebox{90}{Equipartition}}\\
  \includegraphics[width=1.0\linewidth]{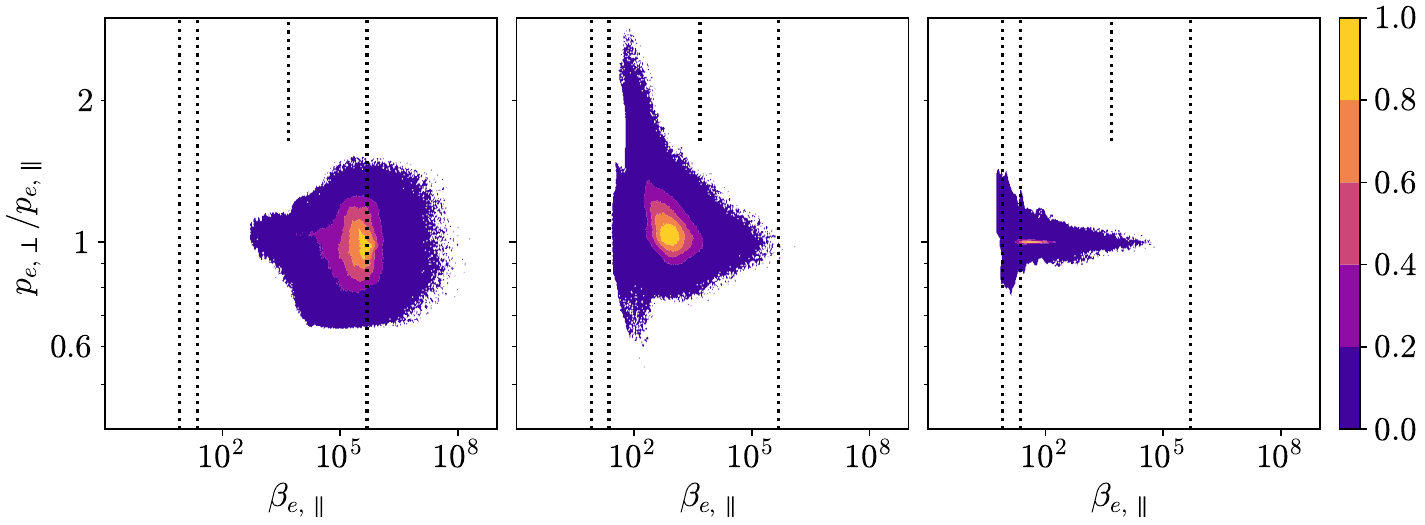}
  \put(-352,30){d)}
  \put(-241,30){e)}
  \put(-130,30){f)}
  \put(-284,30){\tiny \rotatebox{90}{$\rho_e=L$}}
  \put(-313,110){\tiny \rotatebox{90}{$\rho_i=L$}}
  \put(-328,30){\tiny \rotatebox{90}{$\rho_e=k_{\rm Weibel}^{-1}$}}
  \put(-341,30){\tiny \rotatebox{90}{Equipartition}}  
  \caption{\label{fig:brazilk0e1} Normalized probability distributions of $p_{\perp,e}/p_{\|,e}$ and $\beta_{e,\|}$ at three representative time points. Left panels: time of fastest Weibel growth ($t/\tturn=0.038$ and $t/\tturn=0.021$ top and bottom, respectively); middle panels: $t/\tturn=2.0$ (kinematic dynamo phase), right panels: $t/\tturn=10.0$ (saturated dynamo phase). The electron closure parameter is changed from its baseline value $k_{0,e}=1 k_0$ (upper panels) to a negligible value $(2\pi)^{-1}10^{-6} k_0$ (lower panels; note the slightly different y-axis scales). Vertical dotted lines indicate specific magnetic field strength values in terms of $\beta_{e,\|}$. }
\end{figure}

High beta systems where pressure anisotropy driven instabilities are active are often analyzed in terms of their distribution in $p_{\perp,e}/p_{\|,e}$--$\beta_{e,\|}$ space. The distribution may then be effectively limited by stability boundaries \citep{Bale_2009}. Such figures are often referred to as \emph{Brazil plots} due to the characteristic shape of the instability-limited distributions.
We show such distributions in Fig.~\ref{fig:brazilk0e1} at three representative time points in the Weibel, the kinematic, and the saturated phase (the same time points as in Fig.~\ref{fig:cutsL}). The vertical lines indicate magnetic field values for which $\beta_{e,\|}$ is calculated with the initial $p_{e,\|}$. Note that now we show the pressure with components parallel and perpendicular to the magnetic field, even though electrons never get very strongly magnetized. Accordingly, $p_{e,\perp}$ is the average of the generally unequal pressure components perpendicular to the local field direction.     

In the first time point shown (left panels), thermal electrons are just becoming magnetized at the box scale, indicated by the $\rho_e=L$ vertical line. Note that well into the dynamo growth phase, shown by the middle panels, most of the electrons are still far from being magnetized at the spatial scale where the peak of the magnetic spectrum corresponding to the Weibel instability was located (marked by the $\rho_e=k_{\text{Weibel}}^{-1}$ vertical line). That is, the saturation of the Weibel instability is not caused by the electrons becoming magnetized at the characteristic Weibel instability scales. Due to the small pressure anisotropies achieved here, the Weibel saturation mechanism is in its Alfv\'{e}n current regime according to the classification introduced by \citet{Kato_2005}. This limit yields lower saturation field amplitudes, corresponding to $\beta_{e,\|}\approx 1.3\times 10^{3}$ in our simulation, in good agreement with what we observe at the highest $B$ regions upon saturation. 

Since the time evolution of the average ion thermal speed is weak compared to the variation of the magnetic field strength, Fig.~\ref{fig:brazilk0e1} also provides an indication of the evolution of ion magnetization. The vertical dotted lines in the upper part of the panels ($\rho_i=L$) mark the approximate time at which the ions become marginally magnetized on the simulation box scale. Because ions primarily carry the mass flow, it is the flow scale---comparable to the box size---that is the relevant scale for their magnetization. Thus, already at $t/\tturn = 2.0$ (middle panels), the ions may be considered magnetized\footnote{This estimate is based on the root-mean-square magnetic field strength. In principle, if the magnetic field were predominantly at scales much smaller than the corresponding Larmor radius, this measure would not provide a meaningful characterization of magnetization. By this time, however, a sufficiently large fraction of the magnetic energy resides at large scales to justify this interpretation.}. However, even at saturation the magnetic field is not strong enough to strongly magnetize the ions. Moreover, within the simulation domain there are always regions where the magnetic field vanishes, as well as regions of strong field curvature. Accordingly, our use of the term magnetization does not imply complete gyrotropy or conservation of adiabatic invariants. 

At the final time step (right panels), we observe that the $\beta_e$ distribution is effectively limited by equipartition between the energy of the magnetic field and the bulk flows (the latter being calculated at its initial value, which remains representative in the saturated phase).

Notably, as seen by comparing the upper and lower panels, the pressure anisotropies achieved are significantly affected by the closure parameter $k_{0,e}$, somewhat unsurprisingly since the closure drives the system towards isotropic pressure. Without such regularization (lower panels), non-negligible pressure anisotropies can develop during the Weibel instability phase, as the instability has not started to saturate yet; see Fig.~\ref{fig:brazilk0e1}d. 
During the dynamo growth phase, some fraction of the plasma develops even higher anisotropies, especially at lower beta values and in the $p_{e,\perp}/p_{e,\|} > 1$ direction; see Fig.~\ref{fig:brazilk0e1}e. This is consistent with nearly adiabatic electron behavior, where approximate magnetic moment conservation correlates $p_{e,\perp}$ with the overall increase in magnetic field strength. The anisotropy decreases toward higher $\beta_{e,\|}$ values, reminiscent of typical Brazil plots. However, here the significant anisotropy values appear at unusually high (i.e., greater than order-unity) $\beta_{e,\|}$.
This behavior indicates that, in this phase of the simulation, the effective pitch-angle scattering rate caused by pressure anisotropy driven high-beta instabilities is not sufficiently high to pin the anisotropy to stability limits. This behavior is consistent with observations in dynamo simulations performed for pair plasmas kinetically by \citet{Zhou_2024}, and also similar to the kinetic ion dynamics in hybrid simulations of \citet{StOnge_2018}.

In the saturated phase (right panels), the pressure anisotropy is significantly reduced compared to the growth phase, for both $k_{0,e}$ values. This behavior indicates an effective pitch angle scattering, even though most of the electrons are magnetized by the large-scale magnetic fields. The scattering may be due to instabilities, as well as a magnetic field line geometry exhibiting sharp turns acting as scattering centers \cite{Walters_2024,Rosin_2011}.   

\begin{figure}
  \centering
  \includegraphics[width=0.45\linewidth]{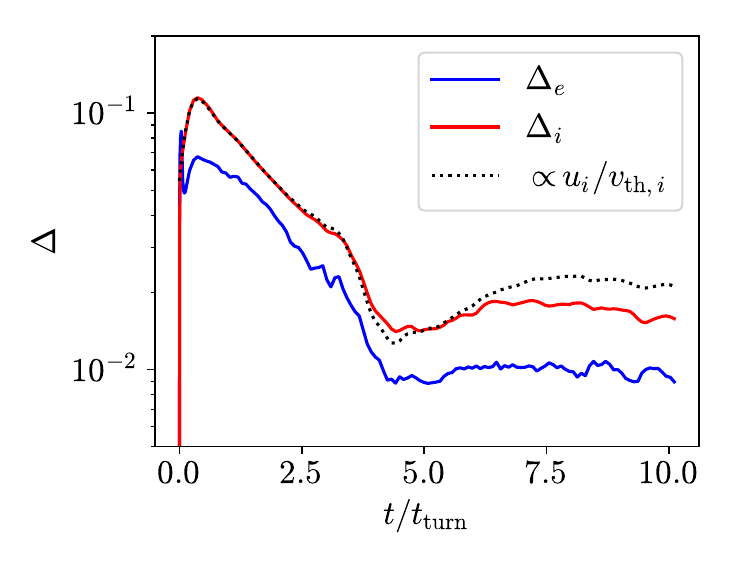}
  \put(-125,115){a)}
  \includegraphics[width=0.45\linewidth]{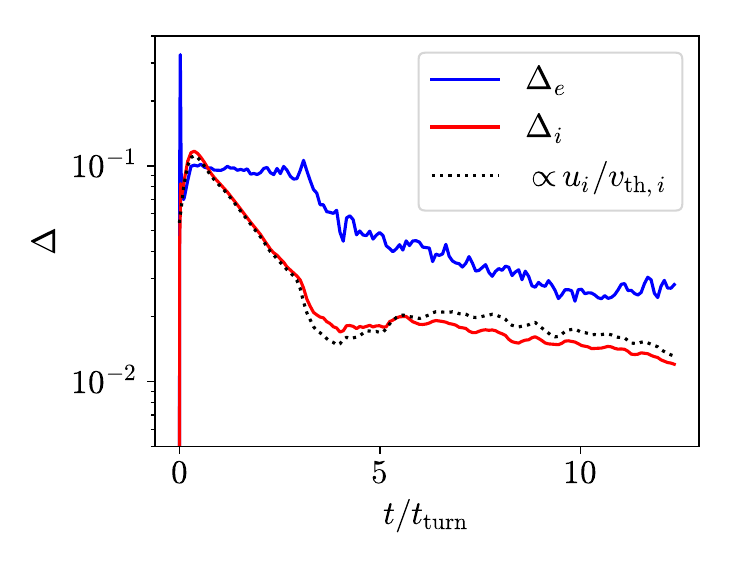}
  \put(-125,115){b)}
  \caption{\label{fig:Deltaslong} Time evolution of the electron (blue) and ion (red) pressure anisotropies for a): $k_{0,e}/k_0 = 1$, and b) $k_{0,e}/k_0=(2\pi)^{-1}10^{-6}\approx 0$. For reference, curves proportional to $u_i/v_{{\rm th},i}$ are also shown with dotted lines.}
\end{figure}

The evolution of the pressure anisotropies during the Weibel phase was presented in Fig.~\ref{fig:baselineShort}. It is also instructive to consider the long-time evolution of these quantities, which is shown in Fig.~\ref{fig:Deltaslong}a for our baseline parameters, in particular $k_{0,e}/k_0 = 1$ and $k_{0,i}/k_0 = 13.6$.
For ions, the observed pressure anisotropy is well described by the same balance used to explain the $k_{0,i}$ dependence of $\gamma_U^{-}$ in Sec.~\ref{sec:reynolds}, namely a competition between production by flow shear and removal by the isotropization closure. This competition leads to the scaling
$\Delta_i \sim k_0 u_i/(k_{0,i} v_{{\rm th},i})$,
where $u_i$ and $v_{{\rm th},i}$ are representative values (e.g., root-mean-square box-averaged quantities). This scaling is supported by the simulation data, as seen from the close agreement between this estimate (black dotted curve) and the measured $\Delta_i$ (solid red curve).

In contrast, no analogous curve is shown for electrons, as $\Delta_e$ is not well described by the same balance. One reason is that the phase-mixing-like production mechanism assumes that the species is not strongly magnetized; indeed, even for ions the agreement deteriorates in the saturated phase, when they become more magnetized. In addition, we employ a weaker isotropization for electrons, so that other processes contribute to the relaxation of pressure anisotropy. Even in the idealized setting of Sec.~\ref{sec:reynolds}, our baseline parameters lie outside the clean $\gamma_B^{-} \propto 1/k_{0,e}$ regime.

In the more complex dynamo simulations, additional processes, such as small-scale magnetic structures, can further reduce $\Delta_e$. This behavior is illustrated by the case where $k_{0,e}$ is essentially zero, while $\Delta_e$ remains regulated, as shown in Fig.~\ref{fig:Deltaslong}b (corresponding to the case in the lower panels of Fig.~\ref{fig:brazilk0e1}). Since electrons are well magnetized on the box scale relevant for the dynamo, pressure anisotropy can in principle be generated by their nearly adiabatic behavior, which, if exact, would lead to an unbounded increase of $\Delta_e$. Instead, once the Weibel phase is over, $\Delta_e$ remains approximately constant during the growth phase of the dynamo and is further reduced in the saturated phase. While this regulation is likely associated with scattering off sharp spatial variations in the magnetic field, a detailed analysis of the underlying mechanisms is left for future work.

\bibliographystyle{jpp}

\bibliography{weibelanddynamo.bib}

\end{document}